\documentclass[pra,amsmath,amssymb,showpacs,floatfix]{revtex4}

\usepackage{graphicx,bm,epsfig,setspace}
\usepackage{color}

\makeatletter
\def\graphicscale{\twocolumn@sw{0.33}{0.4}}
\makeatother

\def\spose#1{\hbox to 0pt{#1\hss}}
\def\lesssim{\mathrel{\spose{\lower 3pt\hbox{$\mathchar"218$}}
 \raise 2.0pt\hbox{$\mathchar"13C$}}}
\def\gtrsim{\mathrel{\spose{\lower 3pt\hbox{$\mathchar"218$}}
 \raise 2.0pt\hbox{$\mathchar"13E$}}}

\def\<{\langle}
\def\>{\rangle}

\newcommand*{\beq}{\begin{eqnarray}}
\newcommand*{\eeq}{\end{eqnarray}}
\newcommand*{\bea}{\begin{eqnarray}}
\newcommand*{\eea}{\end{eqnarray}}

\def\simge{\mathrel{%
       \rlap{\raise 0.511ex \hbox{$>$}}{\lower 0.511ex \hbox{$\sim$}}}}
\def\simle{\mathrel{
       \rlap{\raise 0.511ex \hbox{$<$}}{\lower 0.511ex \hbox{$\sim$}}}}

\begin{document}
\title{Coarse-graining polymer solutions: a critical appraisal of single- and multi-site models}
%\subtitle{Do you have a subtitle?\\ If so, write it here}

\author{Giuseppe D'Adamo}
\email{giuseppe.dadamo@sissa.it}
\affiliation{SISSA, V. Bonomea 265, I-34136 Trieste, Italy}
\author{Roberto Menichetti}
\email{roberto.menichetti@roma1.infn.it}
\affiliation{Dipartimento di Fisica, Sapienza Universit\`a di Roma and
INFN, Sezione di Roma I, P.le Aldo Moro 2, I-00185 Roma, Italy}
\author{Andrea Pelissetto}
\email{andrea.pelissetto@roma1.infn.it}
\affiliation{Dipartimento di Fisica, Sapienza Universit\`a di Roma and
INFN, Sezione di Roma I, P.le Aldo Moro 2, I-00185 Roma, Italy}
\author{Carlo Pierleoni}
\email{carlo.pierleoni@aquila.infn.it}
\affiliation{Dipartimento di Scienze Fisiche e Chimiche, Universit\`a dell'Aquila and
CNISM, UdR dell'Aquila, V. Vetoio 10, Loc.~Coppito, I-67100  L'Aquila, Italy}

\begin{abstract}{
We critically discuss and review the general ideas behind single- and multi-site coarse-grained (CG) models as applied to macromolecular solutions in the dilute and semi-dilute regime. We first consider single-site models with zero-density and density-dependent pair potentials. We highlight advantages and limitations of each option in reproducing the thermodynamic behavior and the large-scale structure of the underlying reference model. As a case study we consider solutions of linear homopolymers in a solvent of variable quality. Secondly, we extend the discussion to multi-component systems presenting, as a test case, results for mixtures of colloids and polymers. Specifically, we found the CG model with zero-density potentials to be unable to predict fluid-fluid demixing in a reasonable range of densities for mixtures of colloids and polymers of equal size. For larger colloids, the polymer volume fractions at which phase separation occurs are largely overestimated. CG models with density-dependent potentials are somewhat less accurate than models with zero-density potentials in reproducing the thermodynamics of the system and, although they presents a phase separation, they significantly underestimate the polymer volume fractions along the binodal. Finally, we discuss a general multi-site strategy, which is thermodynamically consistent and fully transferable with the number of sites, and that allows us to overcome most of the limitations discussed for single-site models.}
\end{abstract}
\maketitle

\section{Introduction}
\label{intro}

Macromolecular fluids are systems characterized by a wide separation of time
and length scales. Length scales range from the local atomic scale 
($\approx$ \AA) to the dimensions of the molecule ($\approx$
1-100 nm), to larger scales, of the order of $\mu$m,
 if mesoscopic particles are included
or the system exhibits self-aggregation into supramolecular structures.
Analogously, time scales vary from 1 ps (local atomic motion) to $\mu$s
and beyond, which 
characterize the conformational relaxation of the chain. Typical 
examples are polymers, either synthetic and/or biological, and 
mixtures with other mesoparticles, like colloids. 
These soft-matter systems show complex physical behaviors, including 
a variety of fluid-fluid and fluid-solid transitions, glassy behavior, 
microphase
separation and supramolecular self-assembly, etc. When focusing on 
these large-scale phenomena, local chemical
details are often of little interest. Therefore, it is useful to develop a
simplified description of the system without including too many
details on the microscopic atomic scale. In this strategy,
commonly referred to as {\it coarse graining}
\cite{Voth2009, PCCP2009, SM2009, FarDisc2010},
one develops models in which most of the internal degrees of freedom 
of the macromolecule are traced out,
projecting the reference macromolecular system onto a 
system with only a limited number of
interaction sites. For colloidal systems, one can use simple
statistical-mechanics models, such as the simple hard-sphere model or
soft-core generalizations thereof. 

A central problem in the coarse-graining approach is the derivation of the
effective interactions among the reduced interaction sites, 
preserving/reproducing at the same time the thermodynamic properties and the 
large-scale features of the underlying microscopic
reference system. For this purpose,
a number of different strategies have been proposed in the
last two/three decades
\cite{Voth2009, PCCP2009, SM2009, FarDisc2010}: 
structure-based methods, energy-based methods,
force-matching methods, relative-entropy methods, to mention the most popular.
In this work we  consider the structure-based route 
to coarse-graining \cite{RS-67,Rowlinson-67,Barker,Casanova,VanderHoef,%
Likos-01,MullerPlathe-02,PK-09},  
with the aim of determining models that accurately reproduce both the 
large-scale and the 
thermodynamic properties of the system under consideration 
in the interesting density ranges, for instance,
in those in which phase separation occurs.
Specifically, we consider single-site models, mapping
each macromolecule on a point-like particle, which, as we will discuss
below, should be appropriate for the investigation of the thermodynamic 
behavior of the system in the low-density regime.  The key
advantage of such an approach is a huge decimation of degrees of freedom and
consequently, a noticeable speed-up of numerical simulations.
Moreover, single-site interaction models can be studied
by using integral-equation methods, which represent
a powerful tool for predicting local structure and 
thermodynamics of simple liquids, whose limitations and validity range  are 
well understood \cite{HMD-06}. 
Similar approaches have also been used for multi-site molecular models,
such as  PRISM (polymer reference interaction-site model) in the 
specific case of macromolecules \cite{SC-87,SC-94,SC-97}.
They are quite successful in concentrated regimes, 
for instance, for polymer melts or polymer
nanocomposites, but not very accurate in the dilute or semidilute 
regime, 
see, for instance, Refs.~\cite{DPP-13-depletion,DPP-14-GFVT} 
for a discussion in the case of polymer-colloid mixtures.
Another advantage of single-site CG models
could be their use in connection with the adaptive strategies recently
developed by Delle Site and coworkers\cite{PDSK-08}; see also Ref.
\cite{PFEDBKED-13} for the recent Hamiltonian formulation of the method.

The first attempt to develop a simple single-site coarse-grained (CG) 
model for macromolecular solutions can be traced back to Ref.~\cite{FK-54},
where two polymer chains with excluded volume at infinite dilution
were considered. Further work on
the issue appeared later in Refs.~\cite{GKK-82,KS-89,DH-94}, with the 
correct determination of the polymer-polymer effective pair
potential. Only recently, however, 
has the single-site CG strategy been employed 
to determine the thermodynamic behavior of these 
complex systems \cite{LLWAJAR-98,WLL-99,LBHM-00,BLHM-01,JDLFL-01,Likos-01}.
The same strategy has also been applied to 
colloid-polymer solutions, representing polymers as monoatomic particles,
interacting by means of a suitable effective potential with the colloids.
This is a generalization of the Asakura-Oosawa model \cite{AO-54}.

In all CG applications, the main issue is the determination of the effective
interactions among the interaction sites at the CG resolution, due to 
their inherently many-body nature.
A possibility, widely used in the literature, is to represent the potential
as a sum of pairwise contributions as derived for vanishingly small
concentration of the macromolecules. Such a potential has a limited range of
validity, being predictive only in the dilute limit, but it has the advantage
of being properly defined, so that the standard statistical-mechanics formalism 
can be used without ambiguities to link thermodynamic properties to 
averages of microscopic observables in the appropriate statistical ensemble.

To extend the validity of the effective interactions in a wider range of
concentrations, while keeping the single-site model, 
two options are possible. One can include progressively
three-body, four-body, etc., terms into the potential when
increasing concentration. Alternatively, 
one can preserve the pairwise structure of
the interaction, but switch to concentration-dependent pair potentials.
The first option provides a systematic method to improve the
transferability of the effective potential in terms of the macromolecular
concentration and preserves thermodynamic consistency.
However, it becomes rapidly unfeasible. Indeed, it 
requires the computation of $n$-body interactions by
performing the statistical average over the internal degrees of freedom 
of $n$ macromolecules while keeping their CG sites fixed in all possible
positions.
The complexity of this task obviously grows exponentially with the
order $n$ and has been attempted only for  the lowest values of
$n$\cite{BLH-01,PH-05,DPP-12-Soft,DPP-12-compressible}.
The second option is apparently easier, since it only requires the
knowledge of suitable two-particle distribution functions as 
in the case of the potentials
derived in the small-concentration limit. However, the pair correlation 
function in a
dense system is not only the result of the direct interaction as derived at
zero density, but it also has a contribution mediated by the other constituents
present in the system. For this reason, the determination
of the effective pair potential from the pair correlation function is 
not straightforward.
One can either use numerical
schemes, such as the iterative Boltzmann inversion 
\cite{RPMP-03} or the inverse Monte Carlo \cite{LL-95,Soper-96}
methods, or 
apply suitable methods within the integral-equation framework of
liquid-state theory.
For instance, in the case of homopolymer solutions 
the hypernetted-chain approximation (HNC) works extremely
well and provides accurate pair potentials \cite{BL-02}.
Unfortunately, in this approach the state dependence of the 
potentials gives rise to several drawbacks and
inconsistencies. For instance, results depend on the 
ensemble one considers \cite{DPP-13-state-dep}. 
Moreover, particular care should be used in deriving the 
thermodynamics, as standard thermodynamic relations do no 
longer hold \cite{Louis-02,DPP-13-state-dep}.

The single-site CG strategy has been applied to 
linear-polymer \cite{LBHM-00,Likos-01,DPP-12-compressible} and star-polymer 
\cite{Likos-01,LLWAJAR-98,WLL-99,JDLFL-01,DLL-02,MP-13,LOB-14} solutions
in the dilute and semidilute 
regime, both in good solvent and in the thermal crossover toward the $\theta$ 
point \cite{DPP-13-thermal}. 
The same strategy has been also applied to study colloid-polymer 
solutions, generalizing the Asakura-Oosawa-Vrij model \cite{AO-54,Vrij-76}.

In this paper we review the single-site CG strategy,
comparing quantitatively results obtained by using zero-density 
and density-dependent potentials. We review the general theory, which is 
then applied to solutions of linear homopolymers and to mixtures of colloids 
and polymers in an implicit solvent. 
We highlight advantages and drawbacks of each option and discuss the 
delicate interplay between state-dependency and ensemble-dependency 
of the interactions, a key ingredient when discussing phase transitions 
as it occurs for instance in the colloid-polymer system. Finally,
we will mention a general multi-site strategy,
which allows us to overcome the limitations of both options,
therefore providing a truly transferrable and consistent CG model for 
polymer solutions in complex situations. 

The paper is organized as follows. In Sec.~\ref{sec:2} we discuss
the general theory behind any structure-based coarse-graining procedure.
Then, as an example, we apply it to the specific case of homopolymer
solutions, discussing in detail the model based on a pairwise zero-density
effective potential. In Sec.~\ref{sec:3} we introduce state-dependent 
pairwise potentials and discuss their limitations in the homopolymer case. 
In Sec.~\ref{sec:4} we generalize these approaches to colloid-polymer 
mixtures and present results for their phase diagram obtained by 
using CG models with zero-density and density-dependent potentials. 
In Sec.~\ref{sec:5} we briefly review a recently developed multisite strategy,
which provides a truly consistent and transferrable CG model, able to 
predict accurately the physics of the underlying microscopic system. 
Finally, in Sec.~\ref{sec:6} we draw our conclusions and perspectives.

\section{Zero-density single-site coarse-grained models}
\label{sec:2}

\subsection{General theory} \label{sec:2.1}

In this section we outline the general theory behind any coarse-graining 
procedure, discussing in detail the different approximations involved. 
An explicit example (homopolymers in implicit solvent) will be discussed 
in Sec.~\ref{sec:2.2}.
Our treatment follows closely Ref.~\cite{Likos-01}. 
In the single-site CG representation, a system of $N$ 
macromolecules 
of $L$ units in a volume $V$ is mapped onto a liquid of point-like particles,
retaining only three translational degrees of freedom (for $d=3$) 
per molecule. In practice, one replaces each macromolecule with a CG particle 
whose position
${\bm R}_\alpha$ is related to the 
positions ${\bm r}_{\alpha,i}$ of the units of the macromolecule by the 
linear transformation
\beq\label{eq:rcm}
{\bm R}_\alpha= {\bm M} ({\bm r}_{\alpha,i}) = \sum_{i=1}^{L} c_{i}{\bm r}_{\alpha,i},
\eeq
where $c_i$ are constant coefficients that identify the representation,
satisfying $\sum_{i=1}^L c_i = 1$.
Typical examples are the center-of-mass (CM) representation, in which 
each macromolecule is represented as a point particle located in
its CM ($c_i=1/L$, for all $i$), and 
the mid-point (MP) representation, in which the 
position of the effective particle coincides with that of 
the central atom ($c_i=\delta_{i,L/2}$). 
In the case of homopolymers, 
the first choice is the usual one for the linear topology
\cite{FK-54,DH-94,LBHM-00,PH-05},
while the second choice is most common when dealing with star polymers 
\cite{WP-86,HG-04,LLWAJAR-98,Pelissetto-12}.

Independently of the details of the representation, in an {\it exact} mapping
the effective potential (free) energy associated with a given CG configuration
$\{{\bm R}_\alpha\}$ can be expressed as \cite{Likos-01}
\beq\label{eq:Veff}
V_{eff}(\{{\bm R}_\alpha\}) =
   V^{(0)}(N,V)+
   V^{(2)}(\{{\bm R}_\alpha \})+
   V^{(3)}(\{{\bm R}_\alpha \})+\dots
\eeq
where $V^{(n)}(\{{\bm R}_\alpha \})$ represents the genuine 
$n$-body contribution, defined as:
\beq
\label{eq:V^n}
V^{(n)}(\{{\bm R}_\alpha\}) =
\sum_{\alpha_1<\alpha_2<\dots<\alpha_n}^{N} 
     u^{(n)}({\bm R}_{\alpha_1},\dots,{\bm R}_{\alpha_n}).
\eeq
The zero-body term, also called {\it volume term}, is independent of the CG 
configuration $\{{\bm R}_\alpha\}$. In the present case, in which 
we only reduce the number of degrees of freedom of the molecule, 
it only contains the free energy of a single isolated macromolecule, hence
$V^{(0)}(N,V) = N v_0$ with no volume dependence.
In the absence of external fields, the one-body term is zero because of the 
translational invariance of the system. Except for the volume term, 
 each $u^{(n)}({\bm R}_{\alpha_1},\dots,{\bm R}_{\alpha_n})$ can be expressed 
recursively in terms of suitable reduced distribution functions in the zero-density limit
 as,
\bea
\label{eq:u^n}
\nonumber
u^{(n)}({\bm R}_{\alpha_1},\dots,{\bm R}_{\alpha_n})&=&
-\frac{1}{\beta}
   \ln\mathcal{G}^{(n)}({\bm R}_{\alpha_1},\dots,{\bm R}_{\alpha_n})-
\sum \left.\! ^{(2,n)} \right. u^{(2)}( {\bm R}_{\alpha_i},{\bm R}_{\alpha_j})\\
\nonumber
&-& \sum \left.\! ^{(3,n)} \right. 
u^{(3)}( {\bm R}_{\alpha_i},{\bm R}_{\alpha_j},{\bm R}_{\alpha_k})\dots \\
&-&\sum  \left. \! ^{(n-1,n)} \right.  
u^{(n-1)}({\bm R}_{\alpha_i},{\bm R}_{\alpha_j},\dots,{\bm R}_{\alpha_l})
\eea
where  $\sum^{(m,n)}$ indicates the sum over all the ordered $m$-ples of $n$ objects, 
and $\mathcal{G}^{(n)}$ is the $n$th-order distribution function
defined as 
\bea\label{eq:Gn}
\nonumber
&\left.  \right. & 
\mathcal{G}^{(n)}({\bm R}_{1},\dots,{\bm R}_{n})=
\langle \exp(-\beta W_{\rm inter}) \rangle_{{\bm R}_{1},\dots,{\bm R}_{n}} \\
\nonumber
&= &\frac{\int  \prod_{\alpha=1}^{n} d{\bm r}_{\alpha}^L e^{-\beta
U_{\rm intra}(\{ {\bm r}_{\alpha,i} \})} e^{-\beta W_{\rm inter}(\{ {\bm r}_{1,i}
\},\dots,\{ {\bm r}_{n,i} \}) }  \prod_{\alpha=1}^n\delta({\bm
R}_\alpha-{\bm M}(\{{\bm r}_{\alpha,i}\}))    } {  \int  \prod_{\alpha=1}^{n}
d{\bm r}_{\alpha}^L e^{-\beta U_{\rm intra}(\{ {\bm r}_{\alpha,i}\} ) } } . \\
&\left.  \right. &
\eea
Here $U_{\rm intra}$ and $W_{\rm inter}$ are the intramolecular and the intermolecular 
interaction potential, respectively. 
For instance, the two-body interaction is given by
\beq\label{eq:u2}
u^{(2)}({\bm R}_1,{\bm R}_2)=
-\frac{1}{\beta}\log {\cal G}^{(2)}({\bm R}_1,{\bm R}_2),
\eeq
which shows that $u^{(2)}({\bm R}_1,{\bm R}_2)$ 
corresponds to the {\it potential of mean force}. 
If the system is homogeneous and isotropic,
this potential depends only on the distance $R=|{\bm R}_1-{\bm R}_2|$,
hence we can simply write it as $u^{(2)}(R)$. 
Analogously, the three-body effective 
potential, defined in Eq.~(\ref{eq:u^n}), is explicitly given by
\bea\label{eq:u3}
\nonumber
u^{(3)}({\bm R}_{1},{\bm R}_{2}, {\bm R}_{3})
&=&-\frac{1}{\beta}\log {\cal G}^{(3)}({\bm R}_{1},{\bm R}_{2},{\bm R}_{3})-
     u^{(2)}(R_{12})-u^{(2)}(R_{13})-u^{(2)}(R_{23})\\
&=&-\frac{1}{\beta} 
\log\frac{{\cal G}^{(3)}({\bm R}_{1},{\bm R}_{2},{\bm R}_{3})
         }{{\cal G}^{(2)}(R_{12}){\cal G}^{(2)}(R_{13}){\cal G}^{(2)}(R_{23})},
\eea
with $R_{ij} = |{\bm R}_i - {\bm R}_j|$.

It is always possible to relate 
the thermodynamics  in the zero-density limit to the effective interactions.
Indeed, by using $u^{(2)}(r)$ one can compute the 
second virial coefficient $B_2$, defined
\cite{Attard-02,HMD-06} by the small-density expansion of the
pressure $P$, $\beta P = \rho (1 + B_2 \rho + \ldots)$,
with $\rho = N/V$, as
\beq\label{eq:B2}
B_2=
-\frac{1}{2}\int d{\bm r} \left(e^{-\beta u^{(2)}(r)}-1\right).
\eeq
Analogous relations hold for the higher-order effective potentials.
For instance, the third virial coefficient $B_3$ can be expressed as 
\cite{DPP-12-compressible}:
\bea\label{eq:B3}\nonumber
B_3&& =-\frac{1}{3}\int d{\bm  r}_{12}d{\bm  r}_{13}
   \left( e^{-\beta u^{(3)}( {\bm  r}_{12},{\bm r}_{13},{\bm  r}_{23})}-1\right)
    e^{-\beta\left[ u^{(2)}(r_{12})+u^{(2)}(r_{13})+u^{(2)}(r_{23})\right]}     \\
&&-\frac{1}{3}\int d{\bm  r}_{12}d{\bm  r}_{13} \left(e^{-\beta u^{(2)}(
r_{12}) }-1 \right)\left(e^{-\beta u^{(2)}( r_{13})} -1\right)\left(e^{-\beta
u^{(2)}( r_{23})}-1 \right).
\eea
These relations show how the thermodynamic and the 
structural properties are intimately related and that a 
structurally consistent mapping 
is {\it necessary} to preserve the correct thermodynamic behavior. 
Therefore, any arbitrary
truncation of the many-body series corresponds to a lack of thermodynamic
consistency between the CG and the original model. These relations also show
that virial coefficients can be used to elucidate the
merits/demerits of any CG model in which one only considers the first few
terms (typically, only the pair potential), as discussed in 
Refs.~\cite{DPP-12-Soft,DPP-12-compressible} for a pure polymeric system and,
in a more general context, in 
Ref.~\cite{AW-14}. 

\subsection{Coarse-graining homopolymer solutions} \label{sec:2.2}

As an example, we apply
the coarse-graining methodology to a solution of homopolymers in implicit solvent.
This is already a CG representation of a real system,
since solvent degrees of freedom have been traced out and replaced by suitable
effective monomer-monomer interactions. In general, their determination
is extremely complex. However, a considerable simplification occurs if one
only considers dilute and semidilute solutions of very long coils---this is the
only case we consider below. In such systems
the local monomer density is very small (it vanishes in the limit in which the
length of the polymers goes to infinity) and the (osmotic) thermodynamic
behavior and the large-scale structure (i.e.,
on scales that are at least of the order of the polymer size)
are universal \cite{deGennes-79,dCJ-book,Schaefer-99}, i.e.,
independent of the microscopic details of the monomer-solvent interactions.
Thus, there is no need to trace out exactly the solvent degrees of freedom.
For instance, to study the good-solvent regime,
it is enough to consider any polymer model, which shows local monomer-monomer
repulsion.

Universality also significantly constrains the effective interactions. 
For finite values of $L$, the distributions $\mathcal{G}^{(n)}$ depend on the 
specific polymer model under consideration, so that also the effective interactions
are model dependent. However, in the scaling limit, i.e., for  $L\to \infty$, 
the adimensional combinations $R_g^{3n}\mathcal{G}^{(n)}$ converge to  
distributions
\beq
 \lim_{L\rightarrow \infty}R_g^{3n}\mathcal{G}^{(n)}({\bm R}_{1},\dots,{\bm R}_{n})=
g^{(n)}({\bm b}_{1},\dots,{\bm b}_{n}),
\eeq
where $R_g$ is the zero-density radius of gyration and 
${\bm b}_\alpha={\bm R}_\alpha/R_g$. 
The distributions $g^{(n)}$ are universal to a large extent as they depend only on the quality of
the solvent. For instance, the same result is obtained by considering any model 
that is appropriate to describe polymer solutions under good-solvent conditions.
Because of the universality of the functions $g^{(n)}$,
also the generic $n$-body effective potential 
$u^{(n)}({\bm b}_{1},{\bm b}_{2},\dots,{\bm b}_{n}$) is universal, i.e.,
independent of the polymer model, once it is expressed in terms 
of the adimensional vectors ${\bm b}_\alpha$,
and uniquely specified once the quality of the 
solvent has been determined.

The first attempts to estimate the low-order terms of the above-reported 
many-body expansion 
can be traced back to the seminal work of Flory and Krigbaum \cite{FK-54}. 
They showed  that, in the CM representation, the effective pair potential 
($n=2$) is approximately Gaussian with a range of the order of the 
radius of gyration of a single chain. Though the functional form of the 
interaction was reasonable, 
their mean-field treatment predicted $u^{(2)}(b=0)$ to scale as $L^{0.2}$
with the length $L$ of the polymer, 
hence it diverged in the scaling, infinite-length limit.
Later, simple scaling arguments \cite{GKK-82},
renormalization-group \cite{KS-89} and numerical \cite{DH-94} 
calculations confirmed the overall shape of the interaction but found that, 
in the scaling limit, the potential is independent of $L$ and it is of the 
order of $k_BT$ at small distances. 
For the MP representation the potential is no longer 
bounded at the origin but diverges logarithmically\footnote{
Witten and Pincus \cite{WP-86} considered star
polymers with $f$ arms and the central-site representation, in which the CG 
molecule is located at the tethering site. For $f=2$ the star polymer is equivalent to 
a linear one, with the CG molecule located at the midpoint monomer.}
as $b\to0$ \cite{WP-86}.

\begin{figure}[tbp]
\begin{center}
\includegraphics[width=0.95\textwidth,keepaspectratio]{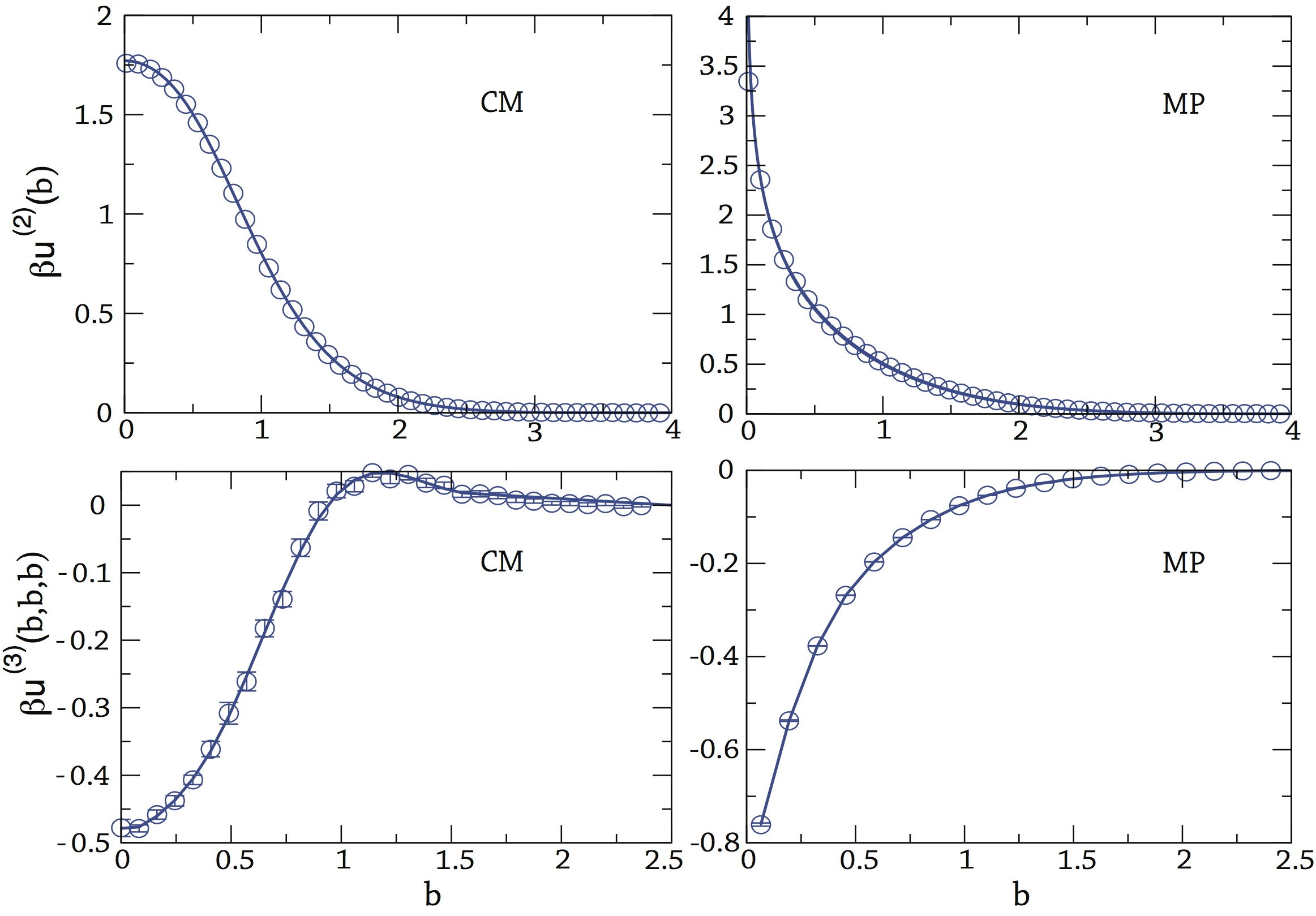}
\caption{Top: Effective pair potential $u^{(2)}(b)$ in 
the CM (left) and MP (right) representations
as a function of $b=r/R_g$. We report (circles) 
numerical DJ results
and (full lines) the corresponding parametrizations 
(\ref{eq:u2cm}) and (\ref{eq:u2mp}).
Bottom: Effective three-body potential 
$\beta u^{(3)}(b,b,b)$ as computed 
in the CM and MP representations as a function of $b=r/R_g$. 
The simulation results (squares)
are obtained by means of Monte Carlo simulations of the DJ model.
The full lines are only meant to guide the eye.
For the CM representation, DJ results agree with those obtained 
by using the SAW model \cite{BLH-01,PH-05}, 
as expected on the basis of universality.
 }
\label{fig:Fig2.5}
\end{center}
\end{figure}

A direct estimate of the effective pair potential can be obtained by
determining the pair distribution function and using 
Eq. (\ref{eq:u2}).
Simulation estimates of the effective pair potential, for both the CM and MP 
representation, are reported in Fig.~\ref{fig:Fig2.5}. We consider 
Domb-Joyce (DJ) 
chains\footnote{At variance with the standard self-avoiding walk 
(SAW) model on the lattice, in which multiple occupancy is strictly forbidden, 
in the DJ model \cite{DJ-72}
self-intersections are possible but with an energy penalty $w$.  
For any positive $w$ such a model describes polymers under good-solvent
conditions in the scaling limit (explicit tests of universality can be 
found in Refs.~\cite{CMP-06,Pelissetto-08,RP-13}). 
The model is extremely convenient for two different reasons. First,
for $w\approx 0.5058$ the leading-order corrections to scaling are absent 
\cite{CMP-06} and 
scaling results are obtained by using relatively short chains
(results for chains with $L=2400$ are effectively scaling-limit 
estimates).
Second, the soft nature of the interactions and the availability of 
very efficient algorithms (for instance, the pivot algorithm
\cite{MS-88,Kennedy-02,Clisby-10}) make simulations very efficient 
(see also Ref.~\cite{Pelissetto-08} for a discussion of simulation algorithms
in the semidilute regime).}
and report results for $L=2400$.
For the CM representation, an accurate parametrization of 
$u^{(2)}(b)$ in the 
scaling limit, was determined by Pelissetto and Hansen \cite{PH-05}, 
using a linear combination of three Gaussian functions:
\beq\label{eq:u2cm}
u_{CM}^{(2)}(b)=\sum_{i=1}^3 a_i \exp(-b^2/c_i^2),
\eeq
with $a_1 = 0.999 225$,	$a_2 = 1.1574$, $a_3 = -0.38505$, 
$c_1 = 1.24051$, $c_2 = 0.85647$, and $c_3 = 0.551876$. 
Such a parametrization was obtained by fitting scaling results obtained 
by extrapolating athermal SAW data. This curve falls on top of the DJ results,
see Fig.~\ref{fig:Fig2.5}, further confirming their universality.
In the MP representation, the potential $u_{MP}^{(2)}(b)$ has been 
discussed at 
length in the context of star polymers. For $b\to0$ it diverges 
logarithmically as $\log(1/b)$ 
\cite{WP-86,vonFerber-etal-2,Pelissetto-12}. 
An explicit parametrization has been given by Hsu {\em et al.}  \cite{HG-04} 
(see their results for a two-arm star polymer)
\beq\label{eq:u2mp}
u_{MP}^{(2)}(b)=
\frac{1}{\tau}\log\left[ 
    \left(\frac{\alpha}{b}\right)^{\tau \beta}\exp(-\delta b^2)+
    \exp(\tau \gamma e^{-\delta b^2}) \right],
\eeq
where $\alpha=1.869$, $\beta=0.815$, $\gamma=0.372$, $\delta=0.405$, 
$\tau=4.5$. A check of these parametrizations can be obtained by 
using Eq.~(\ref{eq:B2}), which implies
\beq\label{eq:A2}
A_2=\frac{B_2}{R_g^3}=
-\frac{1}{2}\int d{\bm b} \left(e^{-\beta u^{(2)}(b)}-1\right).
\eeq
An accurate Monte Carlo estimate of the dimensionless ratio $A_2$ for polymers 
under good-solvent conditions is $5.500(3)$ \cite{CMP-06}. 
If instead we use parametrizations~(\ref{eq:u2cm}) and~(\ref{eq:u2mp}) 
to compute integral~(\ref{eq:A2}) we obtain  $A_2=5.48$ (CM) and 
$A_2=5.51$ (MP), respectively. They are quite close to the direct 
full-monomer (FM)
estimate, confirming the accuracy of the two parametrizations. 

The determination of $u^{(3)}$ requires the computation of a 
triplet correlation, 
a function of three independent scalar variables, which is a cumbersome 
simulation task.  Therefore, we only show results for polymer configurations 
whose CG sites are on the 
vertices of an equilateral triangle of side $b$. The simulations 
results for the 
DJ model are reported in Fig. \ref{fig:Fig2.5}. While in the MP representation 
the effective potential is purely attractive and diverging to $-\infty$ 
as $\rho\to0$ 
(a result that can be proved on general grounds 
\cite{vonFerber-etal-2,Pelissetto-12}), 
in the CM representation it is soft, attractive at short distance ($\rho<1$),
 and with a small repulsive tail for $\rho> 1$.  For the CM representation,
four-body and five-body were also determined \cite{BLH-01}, at least for 
some particular polymer configurations. In particular, at least up to 
$n=5$, it was found that 
the generic $n$-body potential at small distances is positive (repulsive) 
for even $n$'s and negative (attractive) for odd $n$'s. Moreover, 
the strength of the $n$-body potential at small distances decreases for 
increasing $n$. For the MP representation this behavior was proved for all
values of $n$ and generic star polymers with $f$ arms
\cite{Pelissetto-12}: 
moreover, it was shown that the $n$-body potential at full
overlap $b_1 = \ldots = b_n = 0$ decreases logarithmically, at least
for star polymers with a large number of arms.

The full computation of the many-body terms is very difficult and of little
practical use, since numerical simulations of the model using 
three-body or higher-order interactions would be unfeasible. Indeed, the
computational cost grows as $N^n$, where $n$ is the largest
order included and $N$ is the number of constituents of the system.
Therefore, in most of the applications 
the many-body expansion is truncated, considering only
the zero-density pair potential of mean force (\ref{eq:u2}). 
It is therefore important to quantify the accuracy of this very 
simplified effective model.

As a first check of its accuracy,
we consider the universal third-virial combination $A_3 = B_3/R_g^6$, 
which depends explicitly
on the three-body interaction term, see Eq.~(\ref{eq:B3}). Thus, 
comparison of $A_3$ computed in the FM model with the value
computed in the CG model provides us with a direct estimate of the 
quantitative relevance of the neglected three-body interactions.
For polymers in the scaling limit $A_3=9.90(2)$ \cite{CMP-06}. 
Using Eq.~(\ref{eq:B3}) with $u^{(3)} = 0$ and 
parametrizations (\ref{eq:u2cm}) 
and (\ref{eq:u2mp}) we obtain instead
$A_{3,CM}=7.85$  and $A_{3,MP}=4.92$ 
for the two different representations.
The CG model underestimates the 
FM value $A_3=9.90(8)$ in both cases: by $21\%$ in the CM case 
and by $50\%$ in the MP case, respectively.
This shows that three-body interactions are relevant: if they are neglected,
the pressure may be significantly underestimated. Similar conclusions are 
reached by directly comparing the equation of state.
A simple estimate of $Z = \beta \Pi/\rho$, $\rho = N/V$, for the CG system can 
be obtained by using the random-phase approximation (RPA) \cite{HMD-06}, 
which is expected to be accurate for systems with soft potentials
and becomes exact for large densities. If 
$\Phi=(4\pi R_g^3 \rho)/3$, it predicts
\bea\label{eq:EOSMSA}
\nonumber
Z_{RPA,CM}=1+1.71\Phi, \\
Z_{RPA,MP}=1+1.54\Phi.
\eea
Clearly, the CG model does not 
capture the correct scaling of the osmotic pressure in the semidilute regime, 
i.e., $Z \propto \Phi^{1.309}$
\cite{deGennes-79,dCJ-book},  underestimating the correct result.
Moreover, $Z$ depends on the chosen CG representation, a dependence which 
would be absent in the exact mapping.\footnote{Heuristically,
this statement can be understood 
by noting that the isothermal compressibility is 
only related to the fluctuations of the number of polymers, a quantity which 
is invariant under the change of representation.} In particular,
consistently with the results for $A_3$,
the osmotic pressure in the CM representation is always larger than 
the MP representation estimate. 

More quantitatively, we can compare the compressibility factor 
$Z$ in a wide range of densities,
representative of both the dilute and semi-dilute regimes. 
In Table~\ref{Z-table} we report  
FM results \cite{Pelissetto-08} and  estimates obtained
by using the CG models. In the latter case, we show both simulation results 
and estimates obtained by using integral-equation methods with the
HNC closure (they are fully consistent in the whole
density range, confirming the accuracy of the HNC closure for soft potentials). 
The CG model based on the CM representation appears to be more accurate than 
that based on the MP representation. However,
both approaches show significant deviations from the correct FM
estimates in the semidilute regime. This can be easily understood. For 
$\Phi$ larger than 1, polymers overlap, so that
many-body interactions, neglected in the simple CG model with pair potentials,
play a relevant role.

\begin{table}
\caption{Compressibility factor $Z(\Phi) = \beta \Pi/\rho$
for polymers (FM) in the scaling limit \cite{Pelissetto-08} 
and for the CG model in the center-of-mass (CM) and in the mid-point (MP) 
representation.  Both HNC and Monte Carlo (MC) predictions are reported. }
\label{Z-table}
\begin{center}
\begin{tabular}{cccccc}
\hline\noalign{\smallskip}
$\Phi$ & FM & CM-MC & CM-HNC & MP-MC & MP-HNC \\
\noalign{\smallskip}\hline\noalign{\smallskip}
0.135  & 1.187 & 1.18458(1) & 1.185  & 1.17869(1) & 1.182 \\
0.27    &  1.393 & 1.38167(1) & 1.382 & 1.36439(1) & 1.371 \\
0.54	   & 1.854 &  1.80067(1) & 1.803 & 1.74840(1) &  1.762 \\
0.81   & 2.371   & 2.23911(1) & 2.241  & 2.14190(1) & 2.162 \\
1.09	 & 2.959 & 2.70461(1)&  2.707&	2.55534(1) & 2.582 \\
2.18  & 5.634 & 4.55607(2)& 4.559 & 4.18703(1) & 4.240	  \\
4.36	& 12.23 & 8.29709(2)& 8.303 & 7.47886(3) & 7.584 \\
\noalign{\smallskip}\hline
\end{tabular}
\end{center}
\end{table}

The results presented so far are relative to polymers under good-solvent
conditions. 
The CG strategy can be extended to describe solutions in the thermal crossover 
region towards the $\theta$ point, as well \cite{DPP-13-thermal}. 
Indeed, universality holds even in this
intermediate regime if properties are expressed in terms of the 
polymer volume fraction $\Phi$ and of the Zimm-Stockmayer-Fixman \cite{ZSF-53}
$z$-variable, $z\propto (T-T_\theta)L^{1/2}$, which combines the deviation 
of the temperature from the $\theta$ value and the chain length $L$, as 
long as logarithmic corrections (which are relevant at the $\theta$ point)
are neglected \cite{deGennes-79,dCJ-book,Schaefer-99}. Equivalently, and with a
more direct physical meaning, the $z$-variable can be replaced by the 
second-virial combination $A_2$,  whose functional dependence on $z$, 
$A_2(z)$, has been fully characterized in Ref.~\cite{CMP-08}. 
When approaching the good-solvent regime, we have $z\to\infty$ and 
$A_2(\infty)=5.500(3)$, while, when
approaching $T_\theta$, $z\to 0$ and $A_2(z)\approx 4 \pi^{3/2} z$. 
In Ref.~\cite{DPP-13-thermal} 
it has been shown that the CG single-site model with zero-density 
pair potentials becomes accurate in an increasingly 
large density range when approaching the $\theta$ point. 
This can be explained by noting the the $n$-th order virial coefficient
scales as $z^n$ for $z\to 0$, so that
 $n$-body terms become increasingly less relevant 
approaching the $\theta$ point.

The model discussed in Ref.~\cite{DPP-13-thermal} neglects logarithmic 
corrections that scale as $1/\ln L$, hence vanish in the critical limit, 
but which may be relevant for finite values of $L$. Apparently, they
can be neglected when considering the second virial coefficient
or the expansion ratio that gives the variation of the radius of gyration as 
solvent quality changes \cite{PCP-suppl,PIB-suppl,PS-suppl} (see also 
the supplementary material of Ref.~\cite{DPP-14-GFVT} for an extensive 
discussion). However, for some different quantities they are relevant:
for instance, they determine the deviations from the ideal behavior of the 
equation of state at the $\theta$ point in the semidilute  regime and 
the peculiar behavior of the third virial coefficient 
\cite{NNT-91-suppl,ANNT-94-suppl,PH-05}. Unfortunately, it is not possible
to use CG models to account also for these logarithmic corrections.
Indeed, at the $\theta$ point the CG model with only pair potential
is not thermodynamically stable \cite{KHL-03,AJH-04,PH-05}.

\section{State-dependent interactions} \label{sec:3}

\subsection{General theory} 

To enlarge the density range in which CG single-site models with pairwise 
effective interactions can be used, one strategy proposed in the past is 
based on deriving and using state-dependent interactions. Structurally derived
state-dependent potentials have been mostly discussed in the
context of the canonical ensemble, see
Refs.~\cite{SST-02,Louis-02,JHGL-07,DPP-13-state-dep}
and references therein, hence all thermodynamic and structural quantities
depend on the density\footnote{In principle, one should also consider 
the temperature, but such a variable does not play any role in the 
present discussion,
hence it will never be explicitly reported.} $\rho$. In this approach the pair 
potential $u^{(2)}(r;\rho)$ is obtained by requiring the CG model to 
reproduce the two-point correlation function ${\cal G}^{(2)}(r;\rho)$ 
at the given value of the density. The uniqueness 
of such a potential
is guaranteed by Henderson's theorem \cite{Henderson-74,CCL-84}. Of course, 
for $\rho \to 0$ the potential $u^{(2)}(r;\rho)$ converges to the 
potential of mean force $u^{(2)}(r)$ considered before.
The inversion of ${\cal G}^{(2)}(r,\rho)$ to extract 
$u^{(2)}({r};\rho)$ can be
performed by means of iterative procedures. For instance, one could use 
the iterative Boltzmann inversion method
\cite{MullerPlathe-02}. For soft potentials 
the HNC inversion scheme\cite{BL-02,BLHM-01} is also particularly convenient.

Although use of $u^{(2)}(r;\rho)$ allows one to reproduce 
the pair distribution function for any value of the density,
there is no warranty that any other structural property of the underlying 
system---for instance, the three-body correlation function---is reproduced correctly. 
Moreover, state-dependent interactions introduce some inconsistencies
in the calculation of standard thermodynamic properties
\cite{Louis-02,SST-02,DPP-13-state-dep}.  
For instance, in systems with state-independent potentials 
there are two {\em equivalent} routes to the pressure. One can define it 
mechanically (virial pressure),
as the force per unit area acting on the boundaries, or
thermodynamically, as
the derivative of the free energy with respect to density. 
In the presence of state-dependent interactions the two definitions
are no longer equivalent \cite{Louis-02,DPP-13-state-dep}. Moreover, 
in the case of density-dependent potentials 
none of them reproduces the correct pressure of the underlying system,
although, at least in the low-density limit, 
the virial expression is closer to the correct pressure than the thermodynamic
one \cite{DPP-13-state-dep}. 
Another problem of the approach is that effective state-dependent
potentials depend on the ensemble in which they have been derived
\cite{DPP-13-state-dep}: 
the equivalence of the ensembles breaks down.
Therefore, different thermodynamic results are obtained by using 
CG models defined at the same state point of the underlying system
but in different ensembles, 
making the computation of phase transitions and transition lines 
quite challenging. 
As a general message, care is needed when using state-dependent interactions 
to derive the thermodynamics of the original system and to compute free 
energies. In particular, one should be careful to employ state-dependent 
potentials only in the statistical ensemble in which they have been derived.  

\subsection{Homopolymer solutions} \label{sec:3.2}

To elucidate the issues related to the use of state-dependent interactions, 
we consider again
CG single-site models for  linear homopolymers under good-solvent
condition. This case, in the CM representation, 
has been discussed extensively in the past and a 
complete comparison
between the underlying FM system and the CG model, mainly focused on 
thermodynamic, interfacial and large-scale structural properties, has been 
reported  in Refs.~\cite{LBHM-00,BLHM-01,BL-02}. In particular,
Ref.~\cite{BL-02} reports an explicit parametrization of the 
density-dependent pair potential obtained by matching the CM-CM pair
distribution function for SAWs with $L=500$ monomers.
Since interactions are soft
and the CG model corresponds to a monoatomic liquid, the inversion procedure
was performed by using an integral-equation method with the HNC closure. 
This method requires a minimal computational effort and provides accurate 
estimates of the thermodynamic behavior.

\begin{figure}[tbp]
\begin{center}
\includegraphics[width=1.\textwidth,keepaspectratio]{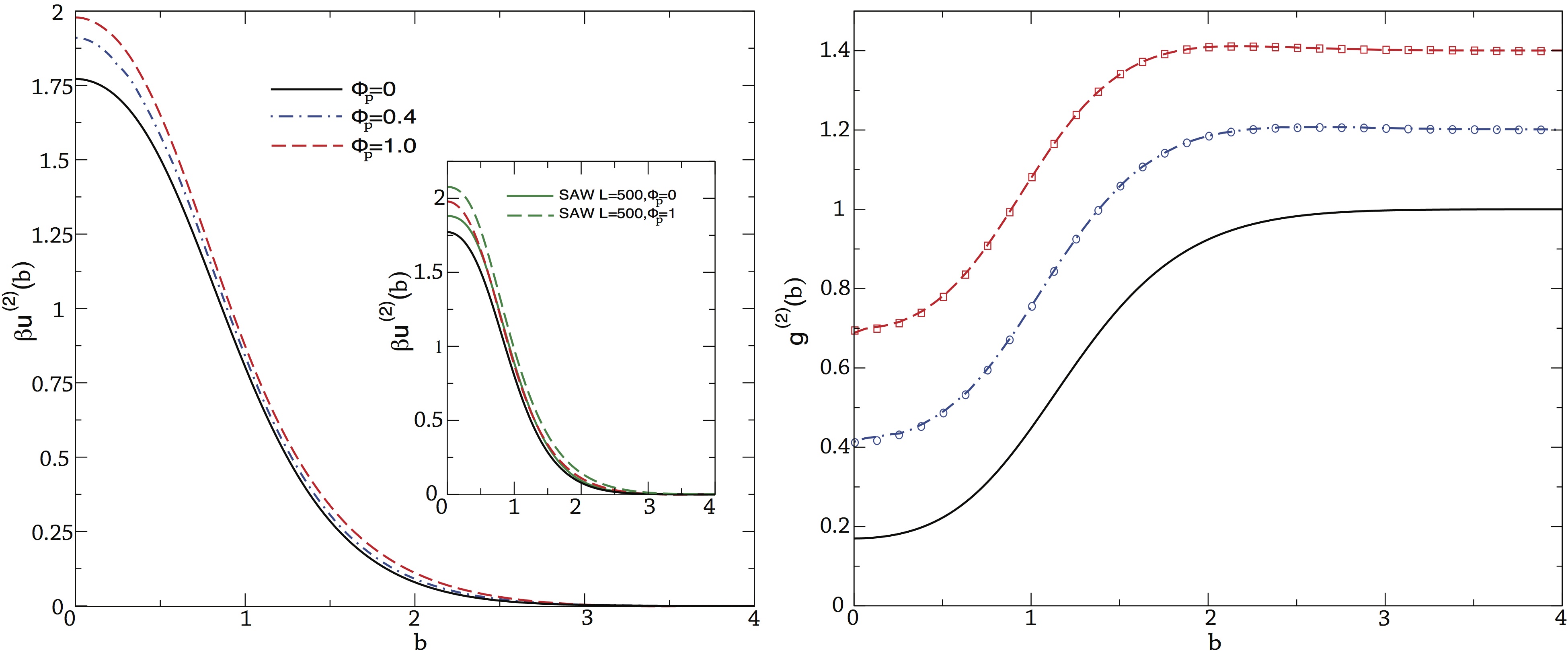}
\end{center}
\caption{Left panel: Effective pair potential $u^{(2)}(b;\Phi)$
for different densities,
$\Phi=0,\Phi=0.4,\Phi=1$, as obtained by HNC inversion
(CM representation). In the inset we compare the effective potentials 
obtained in the scaling limit with those 
appropriate for SAWs with $L=500$ sites.
Right panel: Radial distribution functions between the polymer 
CMs as obtained by Monte Carlo simulations of the  FM model (lines)
and of the CG model with density-dependent potential (squares and circles) for 
$\Phi = 0.4$ and $\Phi = 1$. We also report the zero-density distribution
function (full line). Data for $\Phi = 0.4$ and $\Phi = 1$ are shifted upward
for clarity.
}
\label{fig:Fig2.7}
\end{figure}

In Fig. \ref{fig:Fig2.7} we report the 
effective pair potential $u^{(2)}(b;\Phi)$ (CM representation)
for linear polymers at $\Phi=0,0.4,1$ obtained by using 
the HNC inversion procedure. The associated $g^{(2)}(b;\Phi)$ has been 
computed by FM simulations of the DJ model with polymers of length $L=2400$. 
At first glance, the potentials appear to be not very sensitive to the polymer 
volume fraction. The value at full overlap increases slightly with density
in the range of polymer volume fractions under consideration. 
For larger concentrations 
the strength of the interaction decreases again \cite{LBHM-00,BLHM-01},
as a consequence of the screening of the excluded-volume interaction. 
Moreover, the potential has a slightly longer range compared to the 
zero-density 
case, ensuring the correct scaling behavior of the osmotic pressure in the 
semi-dilute regime. The accuracy of the inversion can be tested by 
performing Monte Carlo simulations of the CG model and comparing the 
resulting pair distribution functions with those used as targets in the 
inversion procedure. From the results shown
in Fig. \ref{fig:Fig2.7} 
we can conclude that the HNC inversion for the CM representation is an 
accurate way to provide structurally consistent effective pair potentials.

The results for the effective potentials reported in Ref.~\cite{BL-02}
differ somewhat from those we have determined by using long DJ chains,
which effectively provide results in the scaling limit. The reason is that 
in Ref.~\cite{BL-02} finite-length SAWs ($L=500$) were considered, 
without performing a scaling-limit
extrapolation. SAW results are affected by relatively large 
scaling corrections, which increase with  
density (for a discussion, see Ref.~\cite{Pelissetto-08}), 
even when $L$ is of order 
$10^3$. In the zero-density limit,
scaling corrections are clearly visible in the result for the pair potential,
which is somewhat more repulsive
than the accurate expression (\ref{eq:u2cm}), obtained by performing 
a proper extrapolation to the limit $L\to\infty$.
In particular, the value at full overlap ($b=0$) of the potential 
exceeds the asymptotic one by $6\%$. A similar
difference is observed for the second virial coefficient,
which takes the value 6.18 if one uses the potential of Ref.~\cite{BL-02}, 
to be compared with the value $5.50(3)$ obtained in the scaling
limit \cite{CMP-06}. To further test the accuracy of the potential,
we have determined the potential at $\Phi=1$ by using the pair distribution
function obtained from
simulations of the DJ model, finding again a discrepancy of approximately 
$6\%$ for the value at full contact, see Fig.~\ref{fig:Fig2.7}. 

Now we analyze the consistency of the results obtained by using 
state-dependent interactions. For this purpose, we consider SAWs with $L=500$ 
as our underlying system, so that we can use the effective density-dependent
potentials reported in Ref.~\cite{BL-02}, which apply to a large $\Phi$ 
interval, up to $\Phi = 2.5$. Then, we 
determine the chemical potential
using three different routes, that are equivalent for systems with 
state-independent interactions. We report results for 
\beq
\beta \hat{\mu} = \ln (\rho R^3_g) + \beta\mu^{(\rm exc)},
\eeq
which differs from the correct 
chemical potential by an irrelevant, model dependent constant, but which has
the advantage of being universal. 
First, we consider  the HNC expression for the chemical potential 
\cite{Morita-60,Attard-91} (HNC-route)
\beq
\beta \hat{\mu}_{\rm HNC}(\Phi)=\ln{\left(\frac{3}{4\pi}\Phi \right )}+
\frac{3}{4\pi}\Phi
\int d^3{\bm b}\left[ h(b)^2-h(b)c(b)-2c(b) \right],
\label{eq:mu_hnc_vir}
\eeq
where $h(b)=g^{(2)}(b;\Phi)-1$ and the direct correlation function $c(b)$ 
is related to $\Phi$ and $h(b)$ by the 
Ornstein-Zernike relation \cite{HMD-06}: 
\beq
h(b_1)=c(b_1)+(3\Phi/4\pi) \int d{\bm b}_2 c(b_2) h(b_{12}),
\eeq
where $b_{12} = |{\bm b}_1 - {\bm b}_2|$.
A second possibility consists in determining first the compressibility 
factor $Z = \beta \Pi/\rho$ by
means of the virial expression,
\beq
Z_{\rm vir} (\Phi) = 
1 - {8\pi^2\over 9\Phi} \int_0^\infty 
    {\partial \beta u^{(2)}(b;\Phi)\over \partial b}
     g^{(2)}(b;\Phi) b^3 db,
\eeq
and then in computing $\hat{\mu}$ as (Z-route)
\bea
\label{eq:mu_compr}
\beta \hat{\mu}_Z(\Phi)&=&\ln\left( \frac{3}{4\pi}\Phi\right)+
Z_{\rm vir}(\Phi)-1+
\int_{0}^{\Phi}\frac{Z_{\rm vir}(\xi)-1}{\xi} d\xi.
\eea
Finally, we consider the compressibility route (K-route),
which is based on 
\beq
\beta \hat{\mu}_{K}(\Phi) =\ln{\left(\frac{3}{4\pi}\Phi \right )}+
\int_0^{\Phi}\frac{K(\xi)-1}{\xi} d\xi,
 \label{eq:mu_hnc_compr}
\eeq
with $K(\Phi)$ given by
\bea
K(\Phi)^{-1}= 
1+\frac{3}{4\pi}\Phi\int d^3{\bm b} \left[ g^{(2)}(b;\Phi)-1\right]. 
\label{eq:compres}
\eea
For CG models with density-dependent interactions, only
the K-route provides the correct chemical potential of the underlying 
model \cite{Louis-02,DPP-13-state-dep}. Indeed, since the CG procedure 
reproduces the pair distribution function at any density, $K(\Phi)$,
defined in Eq.~(\ref{eq:compres}), 
is the same in the CG and in the underlying model. Hence, also 
$\beta \hat{\mu}_{K}(\Phi)$ defined in Eq.~(\ref{eq:mu_hnc_compr}) 
is the same.
Note also that, the Z-route and the K-route 
both require the effective potential to be computed for all densities 
smaller than the 
physical density of interest, hence they have a limited predictive power. 

\begin{table}
\caption{Polymer chemical potential 
$\beta \hat{\mu}$ computed in the density-dependent
CG model appropriate to describe $L=500$ SAWs (we use the 
parametrization of the effective pair potential reported in 
Ref.~\cite{BL-02}). We report results using three different routes 
(the HNC-route, the Z-route, the K-route), as discussed in the text,
and the expression reported in Ref.~\cite{FBD-08} (FBD). The results 
labelled ``FM scaling" are obtained
by using the FM, scaling-limit equation of state reported in 
Ref.~\cite{Pelissetto-08}.
}
\label{table:mu}
\begin{center}
\begin{tabular}{ccccccc}
\hline\noalign{\smallskip}
$\Phi_p$ & HNC-Route & Z-route & K-route & FBD & FM scaling \\
\noalign{\smallskip}\hline\noalign{\smallskip}
0.25 & $-$1.98 & $-$2.02 & $-$2.03 & $-$1.99 &$-$2.11\\
0.5 & $-$0.26 & $-$0.36 & $-$0.43 &$-$0.28 & $-$0.62 \\
1.0 & 3.07 & 2.77 & 2.43 & 3.06 & 1.89 \\
1.5 & 6.83 & 6.14 & 5.41 & 6.80 & 4.35 \\
2. & 11.00 & 9.80 & 8.59 & 10.96 & 6.90 \\
2.5 & 15.28 & 13.73 & 11.98 & 15.42 & 9.55 \\
\hline\noalign{\smallskip}
\end{tabular}
\end{center}
\end{table}

\begin{figure}[htbp]
\begin{center}
\includegraphics[width=0.65\textwidth,keepaspectratio]{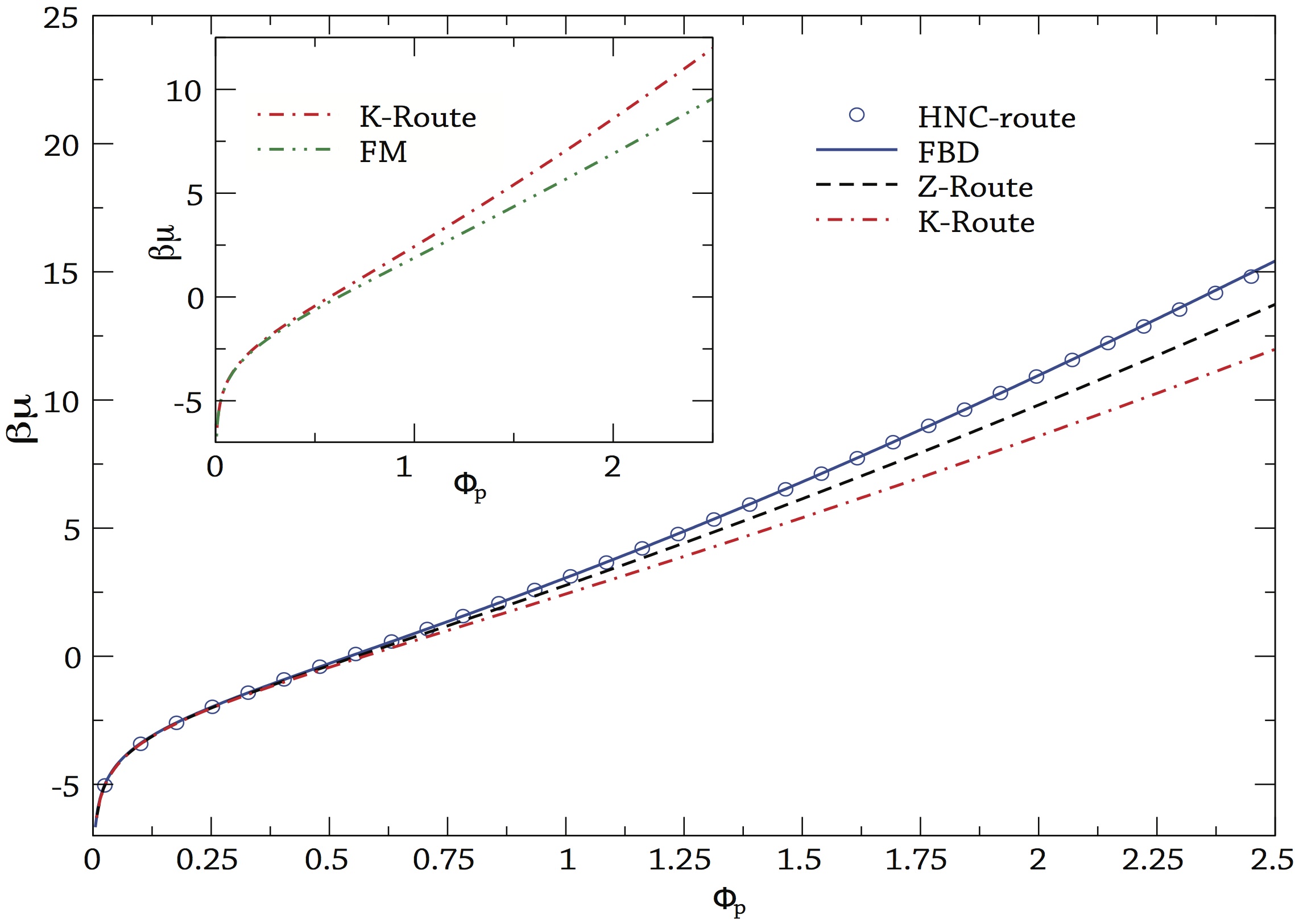}
\end{center}
\caption{Polymer chemical potential $\beta \hat{\mu}$ 
computed in the density-dependent
CG model appropriate to describe $L=500$ SAWs (we use the 
parametrization of the effective pair potential reported in 
Ref.~\cite{BL-02}). We report results using three different routes 
(the HNC-route, the Z-route, the K-route), as discussed in the text,
and the expression reported in Ref.~\cite{FBD-08} (FBD)
In the inset we compare 
the results using the K-route (which are the same as those computed 
directly by using $L=500$ SAWs) and the analogous results obtained 
by using the FM, scaling-limit equation of state reported in 
Ref.~\cite{Pelissetto-08}.
}
\label{fig:mu}
\end{figure}

Results are reported in Table \ref{table:mu} and in Fig.~\ref{fig:mu}. 
It is apparent that inconsistencies between the different routes, 
well beyond the degree of inaccuracy related to the use of the HNC method, 
are present even in the dilute regime. The three different routes provide
different predictions, satisfying
$\beta \hat{\mu}_{\rm HNC}> \beta \hat{\mu}_{Z}>\beta \hat{\mu}_{K}$ for all the
densities under consideration. As a consequence, since the 
K-route estimate agrees with 
the chemical potential of the underlying system, the HNC-route 
and the Z-route both overestimate the correct chemical potential.
It is interesting to observe
that the HNC-route results are equivalent (with a small error
due to HNC approximation) to those that would 
be obtained in a direct 
canonical Monte Carlo simulation 
by employing Widom insertion method \cite{Widom-63}, i.e., this route 
corresponds to the estimate that would be obtained in the 
approach referred to as {\em passive approach} in 
Ref.~\cite{DPP-13-state-dep}. In other words, the HNC-route result is the 
one that would be obtained by using standard thermodynamic relations,
disregarding the density dependence of the potential. As a consequence,
ensemble equivalence is satisfied. If one performs grand-canonical
simulations at chemical potential $\beta \hat{\mu}_{\rm HNC}(\Phi)$ 
with potential $u^{(2)}(b;\Phi)$, one obtains the correct volume 
fraction $\Phi$.\footnote{
Ref.~\cite{FBD-08} checked 
that grand-canonical simulations at $\beta \hat{\mu}$ provide
the correct value of the density, see their Fig.~4. They also provide 
a simple
parametrization of $\beta \hat{\mu}_{\rm HNC} (\Phi)$:
$ \beta \hat{\mu}=
\ln(\rho_p R_g^3)+0.04658+11.05 \rho R_g^3+
35.48 (\rho_p R_g^3)^2-15.71 (\rho_p R_g^3)^3$, 
see Fig.~\ref{fig:mu}. 
}
However, the fact that ensemble equivalence is satisfied is completely
unrelated to the question whether $\beta \hat{\mu}_{\rm HNC}(\Phi)$ is 
a correct estimate of the chemical potential of the underlying system. Indeed,
as our results show,
$\beta \hat{\mu}_{\rm HNC}(\Phi)$ differs significantly from the correct 
result.
Finally, it is interesting to compare $\beta \hat{\mu}_{K}$ obtained here 
(which gives the correct chemical potential for $L=500$ SAWs)
with the chemical potential that is obtained by using 
the equation of state of Ref.~\cite{Pelissetto-08}, which refers 
to polymers in the scaling limit. The two quantities are reported in 
the inset of Fig. \ref{fig:mu}. The SAW model clearly overestimates the 
scaling-limit result, deviations significantly increasing with $\Phi$.

\section{Single-site coarse-grained model for multicomponent systems}
\label{sec:4}

In this section we generalize the discussion of Sec.~\ref{sec:2} 
to multicomponent systems. In particular, we focus on 
colloidal dispersions comprising large particles, colloids, 
usually modeled as hard spheres, and polymers in an implicit solvent. 
These systems are particularly interesting as they show a complex 
phase diagram which depends crucially on the polymer-to-colloid
size ratio: for small ratios, only fluid-solid coexistence is observed,
while for larger values an additional fluid-fluid transition 
is present \cite{Poon-02,FS-02,TRK-03,MvDE-07,FT-08,ME-09}. 
Even in the absence of an explicit solvent, 
the computation of the full phase diagram  is quite difficult,
especially if one is interested in polymers with a large degree
of polymerization. Therefore, 
CG models represent an important tool to investigate these systems.
A first class of CG model is obtained by integrating out all 
polymer degrees of freedom. The resulting CG system is a one-component
model of colloids interacting via an effective potential. Repeating 
the discussion of Sec.~\ref{sec:2.1}, one obtains an effective potential
with an infinite number of many-body terms. Computationally it is 
unfeasible to include more that the leading, two-body term. 
However, such a truncated model is only predictive when the 
polymer-to-colloid size ratio is small. A less extreme approach
consists in integrating out only the internal degrees of freedom 
of the polymer, representing each macromolecule with a monoatomic molecule,
as already discussed in Sec.~\ref{sec:2.2}. After this reduction,
one obtains a two-component system, comprising colloids and 
monoatomic CG polymers, which can be studied with much more
ease than the original system.

Two-component single-site CG models have been considered in 
several papers 
\cite{BLH-02,Dzubiella-etal-01,DLL-02,SDB-03,VJDL-05,PH-06,ZVBHV-09,AP-12} 
and also discussed in Ref.~\cite{DPP-14-GFVT}. Here, we shall only
discuss models in which pair potentials are determined 
accurately by using FM data, in order to assess the 
reliability of the single-site model with pairwise interactions
(other results are summarized and discussed in Ref.~\cite{DPP-14-GFVT}). 
In Refs.~\cite{DPP-13-depletion, DPP-14-colloidi} we performed a 
careful comparison, considering both the model defined at 
zero-density and that using potentials depending on the polymer density
\cite{BLH-02}, focusing on
the solvation properties of a single colloid in a polymer
solution and on the thermodynamics in the homogeneous phase. 
As expected, the model is only accurate if 
$q = R_g/R_c$ is less than 1 
($R_c$ is the radius of the colloid). The failure of the model when 
polymers are larger than colloids can be understood physically,
by noting that, when $q > 1$, 
polymers can wrap around the colloids, a phenomenon that cannot
be modelled correctly if polymers are represented as 
soft spheres.
Moreover, the system is accurate only if the polymer volume 
fraction $\Phi_p = 4 \pi R_g^3 \rho_p/3$ 
(to avoid confusion with colloidal quantities, we add a suffix ``$p$"
to all polymer-related quantities)
is less than 1, guaranteeing that the neglected three-polymer
interactions are small. Finally, the accuracy decreases with
increasing colloid volume fraction $\Phi_c = 4 \pi R^3_c \rho_c/3$
($\rho_c = N_c/V$ is the colloid density), since the relevance
of the polymer-many-colloid interactions increases in this limit. 

Here we discuss the phase diagram of polymer-colloid
solutions as predicted by CG single-site models. To assess their accuracy 
we need reference results to compare with. For $q=1$ we will use 
FM results \cite{CVPR-06,MLP-12,MIP-13}.
To the best of our knowledge,
there are no such results for $q < 1$, hence we will compare 
our Monte Carlo estimates for the CG model with the binodals 
obtained by using the generalized free-volume theory (GFVT)
\cite{FT-08,ATL-02,FT-07,TSPEALF-08,LT-11,DPP-14-GFVT}, which is expected to 
become increasingly accurate as $q$ decreases.

We consider three values of $q$, $q=0.5$, 0.8, and 1.
For the CG models with zero-density and density-dependent interactions,
we perform standard grand-canonical simulations using a recursive 
umbrella-sampling algorithm \cite{TV-77,PRT-14}.
Insertions and deletions of colloids and polymers are performed 
by using the cluster moves introduced by Vink and Horbach 
\cite{VH-04,Vink-04}, which considerably improve the performance
of the simulation. Simulation parameters are the fugacities 
$z_p$ and $z_c$, which are normalized so that 
$\rho_p R_g^3 = z_p$ and
$\rho_c R_c^3 = z_c$ for $\rho_p,\rho_c\to 0$.
Instead of $z_c$ we shall usually quote $\beta \mu_c = \ln z_c$,
while,
as often in the literature, instead of reporting 
$z_p$, we will  report the volume fraction $\Phi_p^{(r)}$
of a polymer reservoir at the same value of $z_p$. For the 
zero-density CG model the  
reservoir volume fraction can be obtained by inverting the 
corresponding equation of state ($z_p = e^{\beta \hat{\mu}_p}$) which we have parametrized as:
\bea
Z(\Phi_p)&=&
{(1+6.05117 \Phi_p+11.6052 \Phi_p^2+ 10.2588 \Phi_p^3)^{1/2} \over
           (1+3.42865 \Phi_p)^{1/2}}\\
\beta \hat{\mu}_p(\Phi_p)&=&\ln\left( \frac{3}{4\pi}\Phi_p\right)+Z(\Phi_p)-1+
     \int_{0}^{\Phi_p}\frac{Z(\xi)-1}{\xi} d\xi.
\eea

\subsection{Results for $q=1$}

\begin{figure}[tbp]
\begin{center}
\includegraphics[width=1.0\textwidth,keepaspectratio]{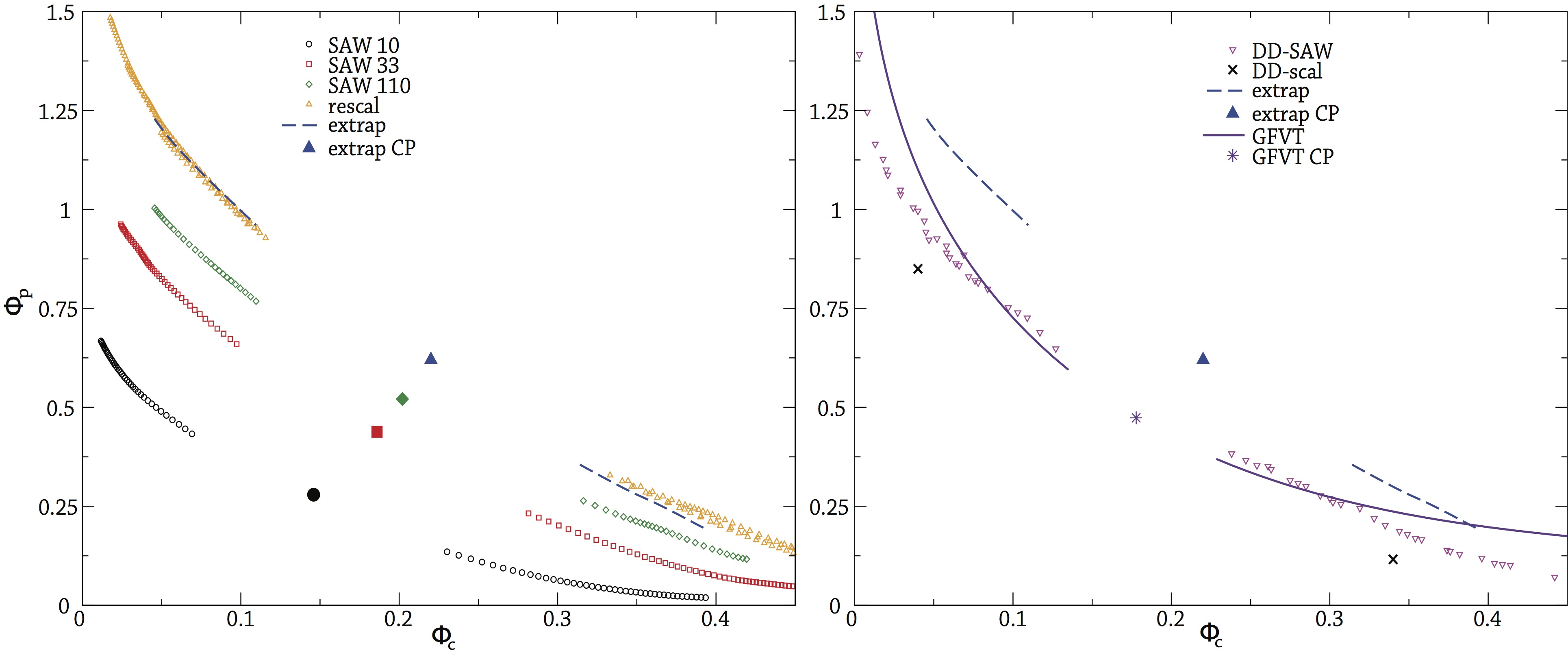}
\caption{Left: binodal curves for $q=1$ obtained by using the 
results of Ref.~\cite{MLP-12}. We report the finite-$L$ 
data ($L=10$, $L=33$, $L=110$), the extrapolation 
obtained by using Eq.~(\ref{binodal-extrapolation}) ("extrap"),
and the binodal obtained by the simple rescaling mentioned in the text
("rescal") starting from the results with $L=110$. 
CP is the extrapolated critical point.
Right: We report the FM binodal (extrap), the GFVT prediction,
and that obtained in Ref.~\cite{BLH-02} using polymer-density-dependent 
potentials appropriate for $L=500$ SAWs (DD-SAW). 
We also report two points (crosses,DD-scal) belonging to the binodal obtained 
using polymer-density-dependent
potentials appropriate for scaling-limit polymers.
We also report the corresponding critical points (CP).
}
\label{fig:Panagio}
\end{center}
\end{figure}

Before studying phase separation by using the CG model, we have determined
the reference binodal, using the FM results of 
Ref.~\cite{MLP-12}. Given the 
computational complexity of the system, the simulated chains are
relatively short. Therefore,
the results of Ref.~\cite{MLP-12} show significant corrections to scaling, 
which should be taken into account before any comparison with the CG results.
The scaling-limit binodal curve can be obtained 
by extrapolating the data of Ref.~\cite{MLP-12}, along the lines 
of the critical-point extrapolation performed in Ref.~\cite{DPP-14-GFVT}.
In Sec.~IV.B of Ref.~\cite{DPP-14-GFVT} we considered the estimates of 
the critical points $\Phi_{c,\rm crit}(L)$ and $\Phi_{p,\rm crit}(L)$,
for three systems with $L=10,33,110$ and approximately $q = 1$, 
and determined the critical point in the scaling limit. We obtained
\cite{DPP-14-GFVT}: $\Phi_{c,\rm crit}(\infty)\approx 0.22$ and 
$\Phi_{p,\rm crit}(\infty)\approx 0.62$. 
Analogously, if $\Phi_p^{\rm bin}(L,\Phi_c)$ gives the position of the
binodal for the system with chains of length $L$, we fit the data to
\footnote{In polymer-colloid mixtures the leading scaling corrections
behave as $L^{-\Delta}$, $\Delta \approx 0.52$ and 
$L^{-\nu}$, $\nu \approx 0.59$, see Ref.~\cite{DPP-13-depletion} for 
a discussion. The two exponents are very close, so that we simply 
extrapolate the data assuming a behavior $a + b/\sqrt{L}$.} 
\beq
\Phi_p^{\rm bin}(L,\Phi_c)\approx \Phi_p^{\rm bin}(\Phi_c)+
\frac{a_1(\Phi_c)}{\sqrt L}.
\label{binodal-extrapolation}
\eeq
The curve $\Phi_p^{\rm bin}(\Phi_c)$ is our estimate of the scaling-limit
binodal.
Another possibility, although less
rigorous, is to rescale, for each value of the length $L$, the 
finite-$L$ binodal 
so as to obtain the correct critical point. In other words,
we set
\beq
\Phi_p^{\rm bin}(\Phi_c) = 
   a \Phi_p^{\rm bin}(L,b \Phi_c)
\eeq
with 
\beq
a = {\Phi_{p,\rm crit}(\infty)\over \Phi_{p,\rm crit}(L)} \qquad
b = {\Phi_{c,\rm crit}(L)\over \Phi_{c,\rm crit}(\infty)}.
\eeq
The binodals computed with this method turn out to be essentially independent
of the value of $L$, supporting the method, and quite close to that
computed by direct extrapolation.
The different extrapolations are reported in
Fig.~\ref{fig:Panagio}, together with the corresponding finite-$L$ results.

Once the reference binodal was determined, we considered the single-site
CG model with zero-density potentials. We systematically increased
$z_p$ and for each value of this parameter we 
performed several runs with different values of $z_c$, covering 
colloid volume fractions from 0.1 to 0.35. In all cases no sign 
of coexistence\footnote{For $\Phi_c = 0.1,0.2,0.3$ the CG model
is homogeneous at least up to 
$\Phi_p = 1.13, 0.82, 0.53$ respectively. 
}
was observed for systems of size $V = (17.7\ R_g)^3$.
Of course, one might fear that systems are too small to allow us to 
identify a phase transition. Therefore, we repeated the analysis
using integral-equation methods. We considered the binary system and used
the HNC closure for all correlations. Again, no sign of phase separation
was observed.

\begin{figure}[tbp]
\begin{center}
\includegraphics[width=0.65\textwidth,keepaspectratio]{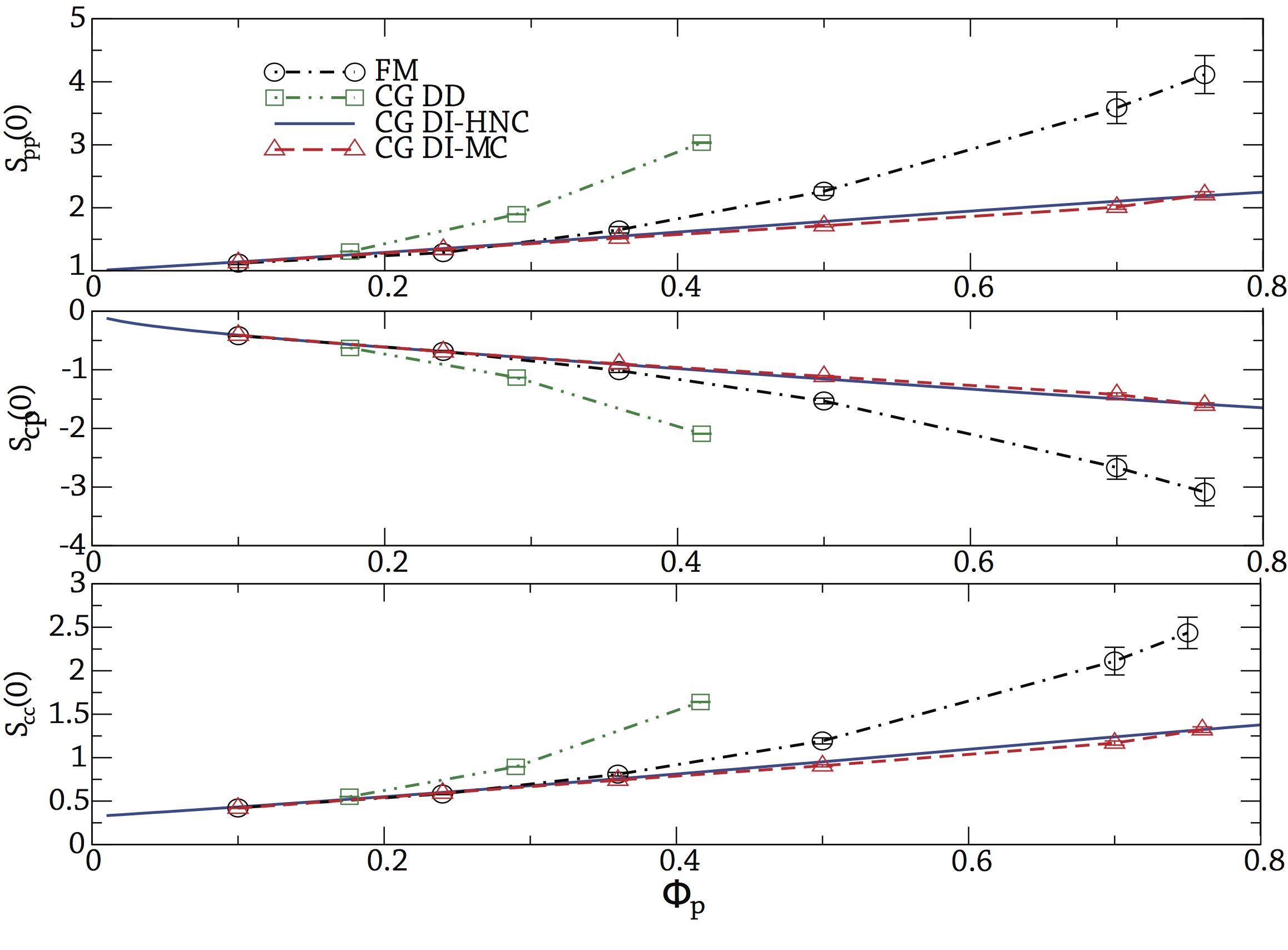}
\caption{Partial structure factors at $k=0$ as a function of 
$\Phi_p$ for $\Phi_c = 0.15$.
We report full-monomer (circles, FM) results obtained by using 
DJ chains of length $L = 2400$, CG results obtained by using 
the zero-density model [Monte Carlo results (triangles, CG DI-MC) and 
HNC results (lines, CG DI-HNC)], and by using the 
density-dependent model (squares, CG DD).
}
\label{fig:SK}
\end{center}
\end{figure}

The evidence of a wide region of stability of the homogeneous 
phase, well beyond the full-monomer phase boundaries, is 
surprising. 
Indeed, one does not expect the single-site model to be accurate 
if colloids and polymers have the same size, hence quantitative 
differences are not surprising. The unexpected feature is 
that the CG model is not even able to
predict the qualitative behavior of the system.

To understand why the CG model does not show phase separation, 
we have determined the partial structure factors $S_{\alpha\beta}(k)$ 
($\alpha,\beta = p,c$) and determined their
limiting value for $k\to 0$. Such quantities are indeed 
order parameters of the fluid-fluid transition. 
We have determined these quantities for
the DJ model with chains of length $L=600$ and for the 
CG model for $\Phi_c = 0.15$
and several values of $\Phi_p$. For the CG model we have determined 
the structure factors both numerically, by performing Monte Carlo 
simulations, and by using integral-equation methods (we use the HNC 
closure) on very large systems $V = (64 R_g)^3$.\footnote{ 
Differences between Monte Carlo and HNC results for the 
CG model increase with increasing 
$\Phi_p$, but are in any case relatively small, confirming the 
accuracy of the final results and the absence of significant finite-volume 
effects. For instance, for $\Phi_p = 1$, well beyond the FM
binodal, we find $S_{pp}(0) = 2.35(8)$ by simulations and 
$S_{pp}(0) = 2.46$ by using the HNC closure. Analogously, we obtain 
$S_{cc}(0) = 1.51(6)$ and 1.62 by using the two different methods.}
Results are reported
in Fig.~\ref{fig:SK}. For small polymer volume fractions, the CG and DJ results are 
in full agreement, but, as $\Phi_p$ increases, the CG model 
significantly underestimates
the structure factors. At coexistence, which should occur for 
$\Phi_p\approx 0.7$-0.8,
the FM estimates are $S_{pp}(0) \approx 4$, $S_{cc}(0) \approx 2.5$,
which are significantly larger than the CG estimates.
More precisely, for $\Phi_p = 0.76$ we obtain
$S_{pp}(0) = 4.1$ and 2.3 for the DJ and the CG model, respectively.
For $S_{cc}(0)$, we obtain correspondingly 
$S_{cc}(0) = 2.2$ (DJ) and 1.3 (CG).
If we further increase $\Phi_p$, the 
CG results change only slightly. We obtain
$S_{pp}(0) = 2.5$ and $S_{cc}(0) = 1.5$ for $\Phi_p = 1.0$.
Clearly, even increasing polymer density the system appears to be unable to 
develop long-range correlations.

The results for the CG model are in contrast with those of Ref.~\cite{BLH-02},
which observed phase coexistence for $q=1$, using the model with 
density-dependent interactions. Quantitatively, the binodal obtained 
in Ref.~\cite{BLH-02} differs somewhat from that obtained by using the 
FM estimates, see Fig.~\ref{fig:Panagio}. 
The results for Ref.~\cite{BLH-02} refer to SAWs with 
$L=500$ monomers, hence one might fear that the differences between
the CG and the full-monomer results are due to the different reference 
system. To clarify the issue, we have redetermined the density-dependent 
potential for $\Phi_p^{(r)} = 1$ using scaling-limit FM
data and recomputed the position of the binodal for such a value 
of $\Phi_p^{(r)}$. We find coexistence between 
$(\Phi_c,\Phi_p) = (0.04,0.86)$ and $(0.34,0.12)$. These two points
are also reported in Fig.~\ref{fig:Panagio}.
They show that the scaling-limit binodal computed by using the 
density-dependent potential is not very different from that 
computed by Ref.~\cite{BLH-02} and still significantly below the 
FM binodal.\footnote{For $\Phi_c = 0.04$, phase 
separation occurs for $\Phi_p = 0.97$ (binodal of 
Ref.~\cite{BLH-02}) and $\Phi_p = 1.24$  
(full-monomer binodal). For $\Phi_c = 0.34$, phase separation 
occurs for $\Phi_p = 0.19$ (Ref.~\cite{BLH-02}) 
and $\Phi_p = 0.30$ (FM).
} 
It is interesting to 
note that the GFVT binodal is essentially on top of the 
binodal of Ref.~\cite{BLH-02}. In view of the previous 
discussion, however, such an agreement looks accidental.

To understand why the CG model with density-dependent potentials
predicts phase separation, we have computed also in this case
the partial structure factors. They are shown 
in Fig.~\ref{fig:SK}. It is clear that the model with 
zero-density potential provides a better approximation to 
the FM results than that using density-dependent potentials. 
However, this is not relevant to obtain phase separation.
The important point is that the CG model with density-dependent potentials
overestimates significantly $S_{pp}(0)$ and $S_{cc}(0)$, hence it 
exhibits phase separation, while the model with zero-density
potentials, although more accurate in the considered range of 
densities, {\em underestimates} $S_{pp}(0)$ and
$S_{cc}(0)$, so that no transition occurs, at least in the range 
we investigated.

\subsection{Results for $q=0.5$ and $q=0.8$}

Let us now consider the behavior for $q=0.5$ and 0.8. In this 
case we do not have reference FM results to compare with. 
Therefore, we use the GFVT predictions that are 
expected to become increasingly accurate as $q$ decreases. Moreover, 
the FM results for $q=1$ provide as an upper bound in $\Phi_p$
on the 
correct binodal. For a given value of $\Phi_c$, phase separation for $q<1$
should occur at polymer volume fractions that are smaller than those 
at which coexistence occurs for $q=1$. We limited our investigation here to the CG model with density independent potential.

\begin{table}[tbp!]
\caption{Binodal line for $q=0.5$. We report the 
values of $\Phi_c$ and $\Phi_p$ in the colloid-gas (g) and 
in the colloid-liquid (l) phase. }
\label{tab:Bino-0p5}
\begin{center}
\begin{tabular}{ccccc}
\hline\noalign{\smallskip}
$\Phi_p^{r}$ & $\Phi_c^{(g)}$ & $\Phi_c^{(l)}$ & $\Phi_p^{(g)}$ & $\Phi_p^{(l)}$ \\
\noalign{\smallskip}\hline\noalign{\smallskip}
0.824 &  0.164  &  0.331  &  0.336  &  0.575  \\
0.827  &  0.157  &  0.340  &  0.325  &  0.588  \\
0.831  &  0.149  &  0.3505  &  0.314  &  0.603  \\
0.8345  &  0.140  &  0.361  &  0.301  &  0.619  \\
0.838  &  0.132  &  0.372  &  0.288  &  0.635  \\
0.842  &  0.123  &  0.382  &  0.275  &  0.651  \\
0.846  &  0.115  &  0.391  &  0.2645  &  0.666  \\
0.8495  &  0.107  &  0.400  &  0.255  &  0.680  \\
0.853  &  0.101  &  0.407  &  0.246  &  0.693  \\
\hline\noalign{\smallskip}
\end{tabular}
\end{center}
\end{table}

\begin{table}[tbp!]
\caption{Binodal line for $q=0.8$. We report the 
values of $\Phi_c$ and $\Phi_p$ in the colloid-gas (g) and 
in the colloid-liquid (l) phase. }
\label{tab:Bino-0p8}
\begin{center}
\begin{tabular}{ccccc}
\hline\noalign{\smallskip}
$\Phi_p^{r}$ & $\Phi_c^{(g)}$ & $\Phi_c^{(l)}$ & $\Phi_p^{(g)}$ & $\Phi_p^{(l)}$ \\
\noalign{\smallskip}\hline\noalign{\smallskip}
1.604 & 0.177 & 0.327 & 0.666 & 1.09  \\
1.610 & 0.175 & 0.330 & 0.661 & 1.10  \\
1.616 & 0.173 & 0.333 & 0.655 & 1.115  \\
1.621 & 0.171 & 0.337 & 0.649 & 1.125  \\
1.627 & 0.168 & 0.340 & 0.643 & 1.14  \\
1.633 & 0.166 & 0.344 & 0.637 & 1.15  \\
1.639& 0.163 & 0.347 & 0.631 & 1.16  \\
1.645 & 0.160 & 0.351 & 0.624 & 1.18  \\
1.650 & 0.157 & 0.354 & 0.617 & 1.19  \\
1.656 & 0.154 & 0.358 & 0.610 & 1.20  \\
1.662 & 0.151 & 0.362 & 0.604 & 1.22  \\
1.668 & 0.148 & 0.365 & 0.597 & 1.23  \\
1.67346 & 0.145 & 0.369 & 0.590 & 1.24 \\
\hline\noalign{\smallskip}
\end{tabular}
\end{center}
\end{table}

To identify the coexistence line, we proceed as follows. 
We fix $z_p$ and determine the distribution of $N_c$ and $N_p$
for several values of $z_c$, either directly or 
by applying the standard reweighting method \cite{FS-89}. 
Then, the value of $z_c$ corresponding 
to the binodal, $z_{c}^{\rm bin}(z_p)$, is obtained by applying the usual 
equal-area criterion: the areas below the two peaks characterizing 
the distributions of both $N_c$ and $N_p$ should be equal. 
Once $z_{c}^{\rm bin}(z_p)$ has been identified, the averages
of $N_c$ and $N_p$ over the two peaks give the number of 
polymers and colloids in the two phases. Results are reported in
Tables~\ref{tab:Bino-0p5} and \ref{tab:Bino-0p8} for $q=0.5$ and 
$q=0.8$, respectively. They have been obtained using reasonably large cubic systems,
of side $31.2 R_g$ and $23.1 R_g$ for $q=0.5$ and $q=0.8$, respectively. 
We expect size effects to be negligible, except possibly close to the 
critical point.

\begin{figure}[t]
\begin{center}
\begin{tabular}{c}
\includegraphics[width=0.5\textwidth,keepaspectratio]{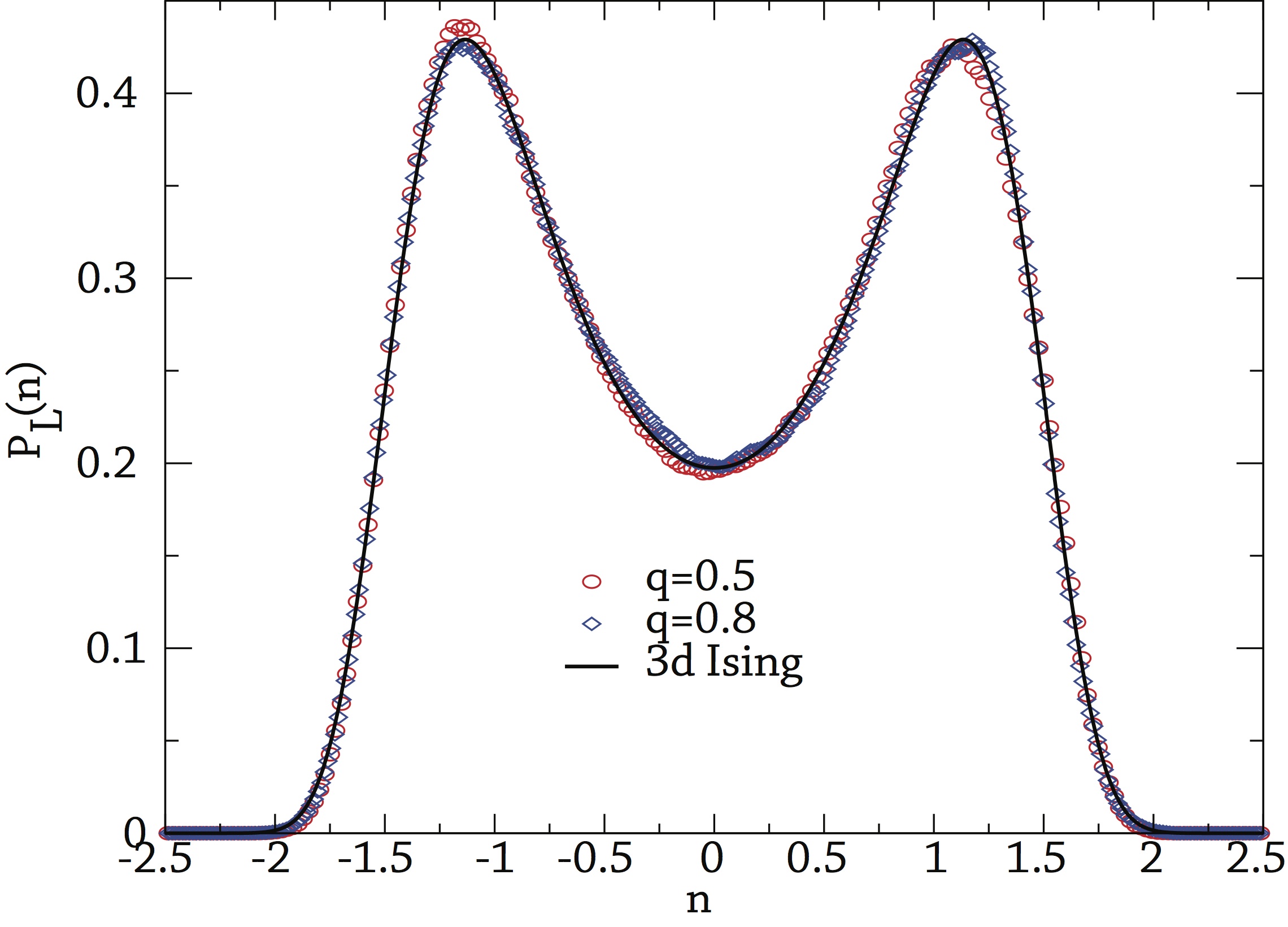} \\
\end{tabular}
\end{center}
\caption{Critical point distribution. 
The abscissa is rescaled to obtain a unit variance distribution.
}
\label{fig:ISINGB}
\end{figure}

To identify the critical point we use the method of Wilding 
\cite{Wilding-97}, exploiting the fact that the transition is in the same 
universality class as the three-dimensional Ising transition. 
In the spin system
the order parameter is the magnetization $M$, whose distribution
at the critical point is known quite accurately
\cite{TB-00}:
\beq
P(M) =  A \exp\left[ 
    - \left({M^2\over M_0^2} - 1\right)^2 \left(c + a {M^2\over M_0^2}\right)
     \right],
\label{Ising}
\eeq
with $A = 0.486642$, $a = 0.158$, $c = 0.776$. The normalization constant
$M_0^2$ can be determined by noting that $\langle M^2 \rangle = 0.777403
M_0^2$. For the mixture the order parameter analogous to the magnetization is 
a linear combination of $N_c$ and $N_p$ that can be defined as 
$n =  A (N_c - a N_p + b)$.  
Then, using the distributions of $N_c$ and $N_p$ computed for
each value of $z_p$ and $z_c^{\rm bin}(z_p)$, we determine 
$a$ and $b$ by requiring $\langle n \rangle = 0$ and the 
distribution to be symmetric around $n = 0$. Finally, $A$ is 
determined by requiring $\langle n^2 \rangle = 1$. Thus, for each value
of the binodal we obtain a distribution function of the variable $n$, 
which is compared with distribution (\ref{Ising}). The best matching 
occurs at a value $z_p$, which is then identified with the critical point.
The distributions at the critical point are compared with 
the Ising one in Fig.~\ref{fig:ISINGB},
where we report the Ising distribution with $n = 1.13417 M/M_0$ and the 
distributions obtained using the data for $q =0.5$ and 0.8. The 
agreement is very good. The mixing of $N_c$ and $N_p$ is very small. We obtain
$a = 0.069$, and 0.148 for $q = 0.5$ and $q = 0.8$,
respectively. This is further confirmed by the 
distributions of $N_c$ shown in Fig.~\ref{fig:PPhic}: they are already quite 
symmetric along the binodal.

\begin{figure}[t]
\begin{center}
\begin{tabular}{c}
\includegraphics[width=0.65\textwidth,keepaspectratio]{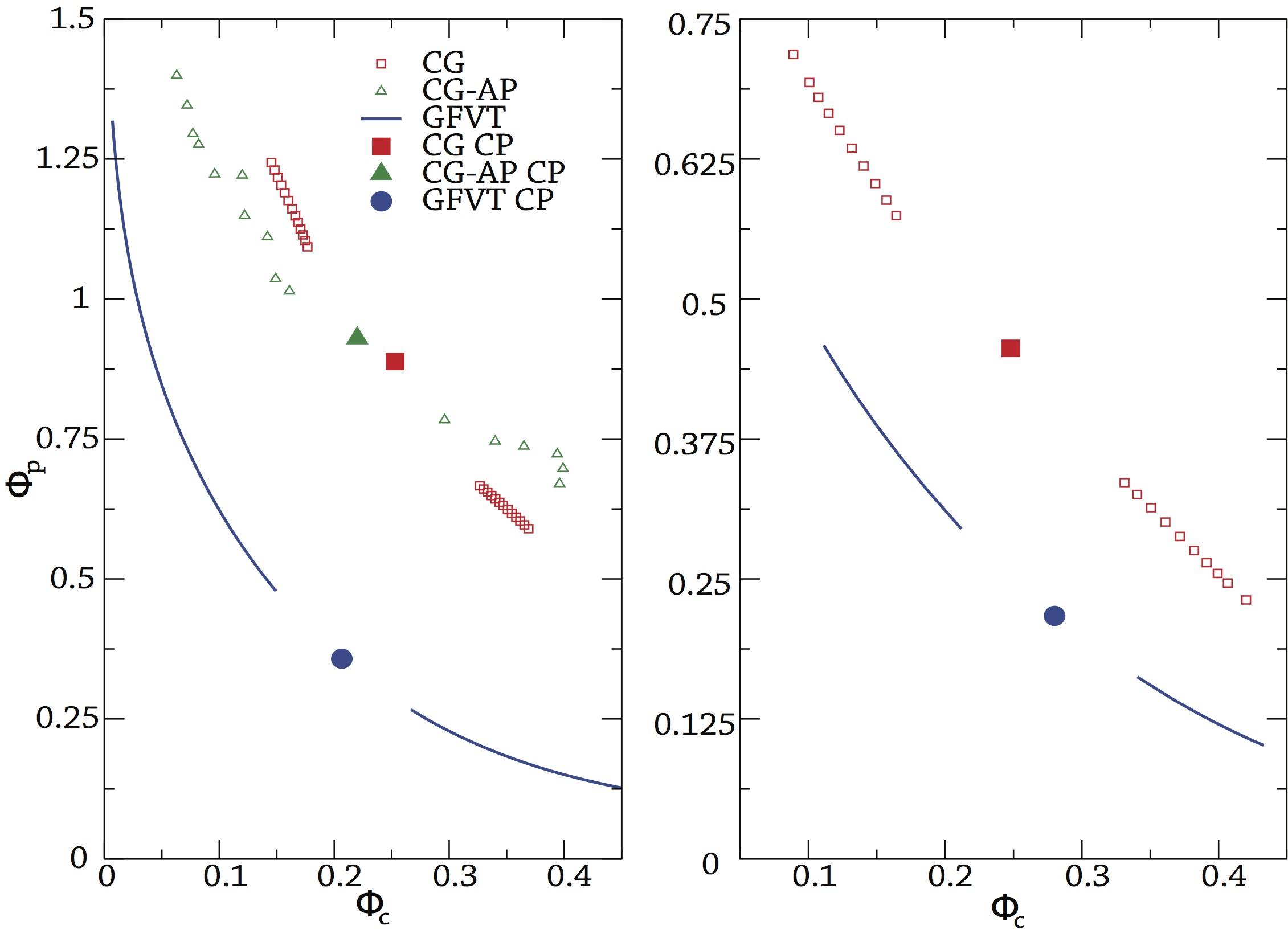} \\
\end{tabular}
\end{center}
\caption{Fluid-fluid binodals for $q=0.8$ (left) and $q=0.5$ (right).
We report the single-site (CG) result and the GFVT prediction.
For $q=0.8$ we also report the binodal computed using the simplified 
model of Ref.~\cite{AP-12} (CG-AP). For each binodal we also report the 
corresponding critical point (CP).
}
\label{fig:binodal-q0p8-q0p5}
\end{figure}

For $q = 0.5$, the analysis of the data 
gives $z_{p,\rm crit} = 2.28$ (equivalently
$\Phi_{p,\rm crit}^{(r)}=0.823$), $\beta \mu_{c,\rm crit} = 27.2$. 
Correspondingly, we have $\Phi_{c,\rm crit}=0.25$ and 
$\Phi_{p,\rm crit} =0.46$.
We have not performed a detailed analysis of the finite-box error
on these results, but it should be of the order of 0.01 on both 
critical volume fractions.

For $q = 0.8$ the analysis of the data gives $z_{p,\rm crit} = 60.11$ 
(equivalently $\Phi_{p,\rm crit}^{(r)}=1.621$)
and $\beta \mu_{c,\rm crit} = 22.9$. Correspondingly,
we obtain $\Phi_{c,\rm crit}=0.25$ and $\Phi_{p,\rm crit} =0.89$.

Let us now compare the results with other estimates. For $q = 0.8$ it is quite 
evident that the single-site binodal is located at polymer densities
that are too large. This is quite evident if we consider the location 
of the critical point. For $q=1$ we estimated $\Phi_{c,\rm crit} = 0.22$ 
and $\Phi_{p,\rm crit} \approx 0.62$ for the full-monomer model 
\cite{DPP-14-GFVT}, hence the obtained estimate of $\Phi_{p,\rm crit} =0.89$
is clearly far too large. It is interesting to note that 
the simplified model of Ref.~\cite{AP-12} gives a binodal which is not 
very different from the one computed here. 
For $q = 0.5$ the single-site binodal is compatible with the upper bound 
provided by the full-monomer results with $q= 1$. However, comparison
with the GFVT results indicate that, most likely, the single-site CG model
predicts phase separation at values of $\Phi_p$ that are too large.

\begin{figure}[t]
\begin{center}
\begin{tabular}{c}
\includegraphics[width=0.65\textwidth,keepaspectratio]{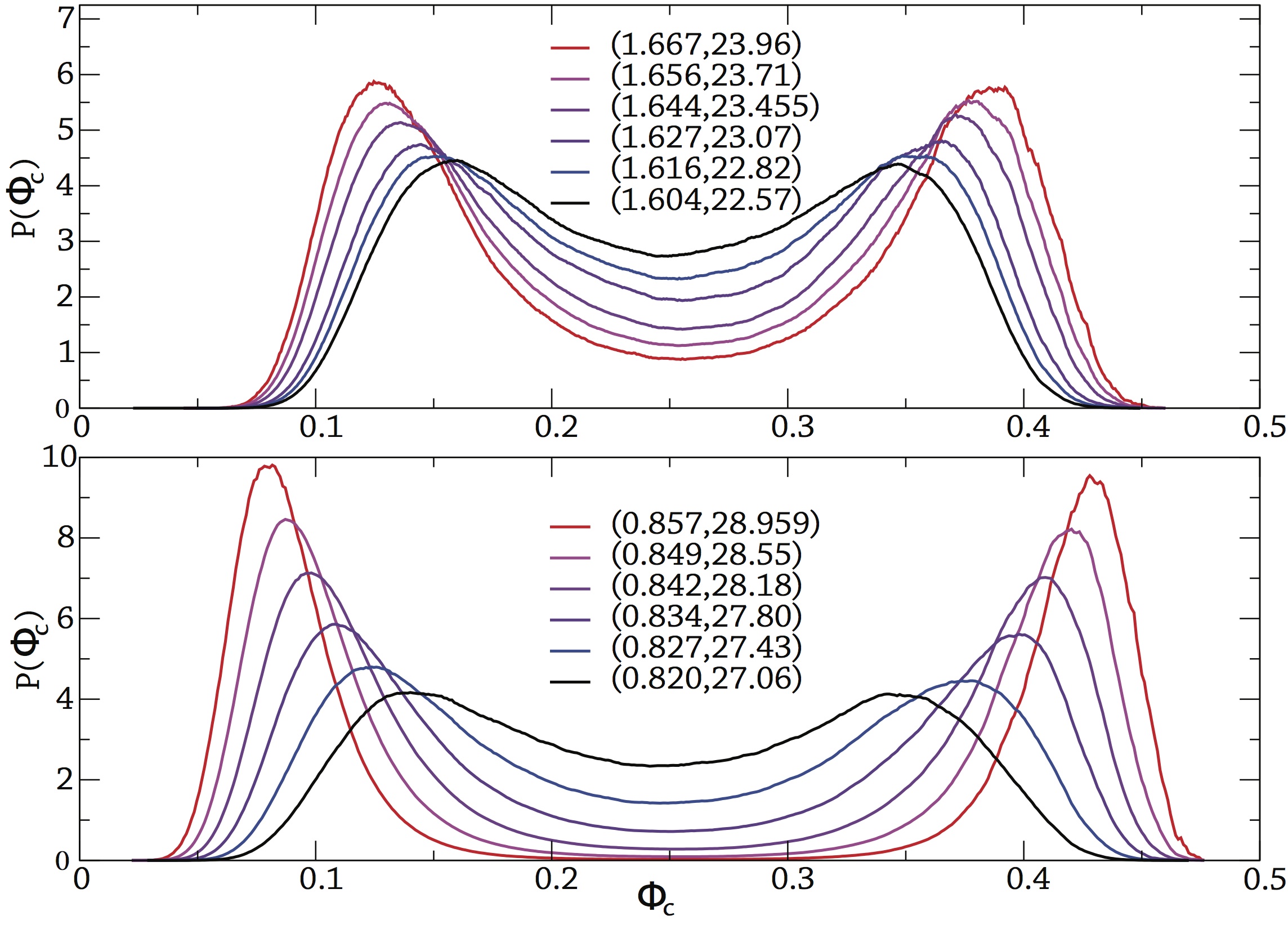} \\
\end{tabular}
\end{center}
\caption{Distribution of $\Phi_c$ at the binodal: (top) $q = 0.8$, 
(bottom) $q=0.5$. We report curves for different values of 
$\Phi_p^{(r)},\beta \hat{\mu}_c$ (they are reported in the legend).}
\label{fig:PPhic}
\end{figure}

\section{Multi-site coarse-grained models} \label{sec:5}

In the previous sections we have discussed single-site CG models 
and shown their advantages and limitations in describing the physics of 
the underlying microscopic system. Models with zero-density potentials 
are thermodynamically and structurally consistent, 
but are only accurate in a
narrow range of densities (dilute regime) 
and, in the presence of colloids, for 
quite small polymer-to-colloid size ratios $q$. For instance, for $q=1$ 
we have not been able to observe phase demixing 
in a reasonable range of densities, 
while for smaller values of $q$ 
the polymer densities at which demixing occurs are significantly 
overestimated.
The single-site model with state-dependent potential are only apparently
more promising. They are not more accurate than those using 
zero-density potentials and moreover, 
as discussed in Sec.~\ref{sec:3}, they are not thermodynamically
consistent.

It seems therefore quite difficult to devise an accurate and consistent 
CG model at the level of the single-site representation with pairwise 
interactions. On the other hand, 
the route of single-site models with many-body, state-independent 
interactions is also impractical. 
Hence, it is tempting to abandon the single-site models in favor of the 
multi-site representation. For polymeric chains in the dilute and semidilute
regime, multi-site representations allow us to extend the 
density range in which they are predictive,
because of the fractal nature of the chains. The volume occupied by a 
long chain in solution scales like $R_g^3\sim L^{3\nu}$ where $\nu>1/3$ is the 
Flory scaling exponent 
($\nu=0.5$ in $\theta$ solvent and $\nu\simeq 0.5876$ in good solvent). 
Moreover, since chains are self-similar objects, the volume occupied 
by each section of a chain of $\ell$ monomers (blob, $\ell\gg 1$) scales 
like $R_b^3(\ell)\sim \ell^{3\nu}$ where $R_b$ is the radius of gyration of 
the polymer section. We have seen that the single-site CG model with pair 
interactions derived at zero polymer density provides an accurate representation of polymer solutions as far as $\Phi_p=4\pi R_g^3 N/3V$ 
is at most 1. Let us now divide each chain in $n$ blobs of size $\ell$
 in such a way that $L=n\ell$ and let us represent each polymer section 
by a single interaction site. By using state-independent interactions
 we can expect this model to be accurate as far as 
$\Phi_b=4\pi R_b^3(\ell)nN/3V=\Phi_p n(\ell/L)^{3\nu}=\Phi_p n^{1-3\nu}\leq 1$,
i.e., for $\Phi_p\leq n^{3\nu-1}$.
Since $3\nu-1>0$, this relations implies that 
we can increase the density range of validity of 
the multi-site model and explore the semidilute regime 
by increasing the number of blobs per chain. 
Since interactions are derived at zero density
these models preserve thermodynamic consistency.

These ideas were first used in constructing simple multiblob models based on pair-wise intermolecular blob interactions, simple bonding intramolecular interactions and simple transferability assumptions. In Ref.~\cite{PCH-07-multi} 
potentials obtained at zero-density were employed while in 
Ref.~\cite{P-09-multi} the potentials were optimized to reproduce properties at finite density from full monomer simulations. 
From these first studies it was clear that multi-site models are more difficult than their single-site analogs. 
The additional difficulty comes from the fact that intramolecular and 
intermolecular interactions 
have intrinsically a many-body character. 
When $\Phi_p < n^{3 \nu - 1}$ we can safely neglect
the interactions among three or more polymer chains.
However, the interactions
among blobs belonging to the same molecule (intramolecular
interaction) or to different 
molecules (intermolecular interaction) are not necessarily well represented as sums 
of pairwise potentials. Indeed, there is no 
obvious "small" physical parameter that allows us to expand the
many-body blob potentials as a sum of two-body, three-body, $\ldots$ terms.
At small polymer density ($\Phi_p<1$), one could expect such a parameter 
to be the ratio between the volume of a blob and the volume of the 
chain $(R_b/R_g)^3=n^{-3\nu}$. This will indeed asymptotically control 
the importance of two-body nonbonding interactions relative to the 
many-body nonbonding terms. However,
this parameter does not take into account the connectivity of the chain 
and its topology.
It should be emphasized that the nature of the CG strategy is to replace long polymers by CG models with a limited number of sites. 
Therefore, we would like to avoid increasing the number of blobs per chain 
too much, in particular, we would like to use this parameter to control the 
accuracy in polymer density, not the accuracy of the single-chain properties. 
At the same time, a good CG strategy should be able to reproduce the 
structural properties of a single chain (related to the intramolecular effective potential) with any number of sites in a reasonable range of $n$.

In Ref.~\cite{DPP-12-Soft} the first step toward such universal multi-site 
CG representation of polymer solutions was presented. We proposed to map 
a single linear chain in the scaling limit, the properties of which, suitably 
normalized, are universal properties of self-avoiding paths, onto a tetramer, 
i.e., a linear molecule with four interaction sites. The intramolecular force
field of the tetramer was decomposed in two-body, three-body and four-body terms inspired by  the way used in building the force fields of real molecules. 
Namely we introduced pair interactions between any pair of sites, 
three-body interactions in terms of bending angle potentials, and four-body interactions in terms of a dihedral angle potential. 
The potentials were obtained numerically by iteratively inverting the
appropriate reduced probability densities. Details are given in
Ref.\cite{DPP-12-Soft}. The choice of a tetramer representation over more
general $n$-mer representations with $n>4$ was dictated by simplicity reasons:
the tetramer is the most elaborated representation in which all many-body
interactions (up to four body) can be parametrized as scalar functions of
suitably defined one-dimensional variables. In a pentamer, the coupling between
the two dihedral angles requires a scalar function of two scalar variables
which is more difficult to reproduce. On the other hand, we did not limit ourself to dimer or trimer representations since we could not find a suitable 
transferability assumption allowing us to use these models as building blocks 
for more refined CG models (tetramer or more blobs) while keeping the desired
accuracy. The tetramer CG model was found to reproduce full monomer results for
polymer solutions 
under good solvent conditions up to $\Phi_p\simeq 2$ with a $\sim 5\%$ accuracy.

\begin{figure}[t]
\begin{center}
\begin{tabular}{c}
\includegraphics[width=0.65\textwidth,keepaspectratio]{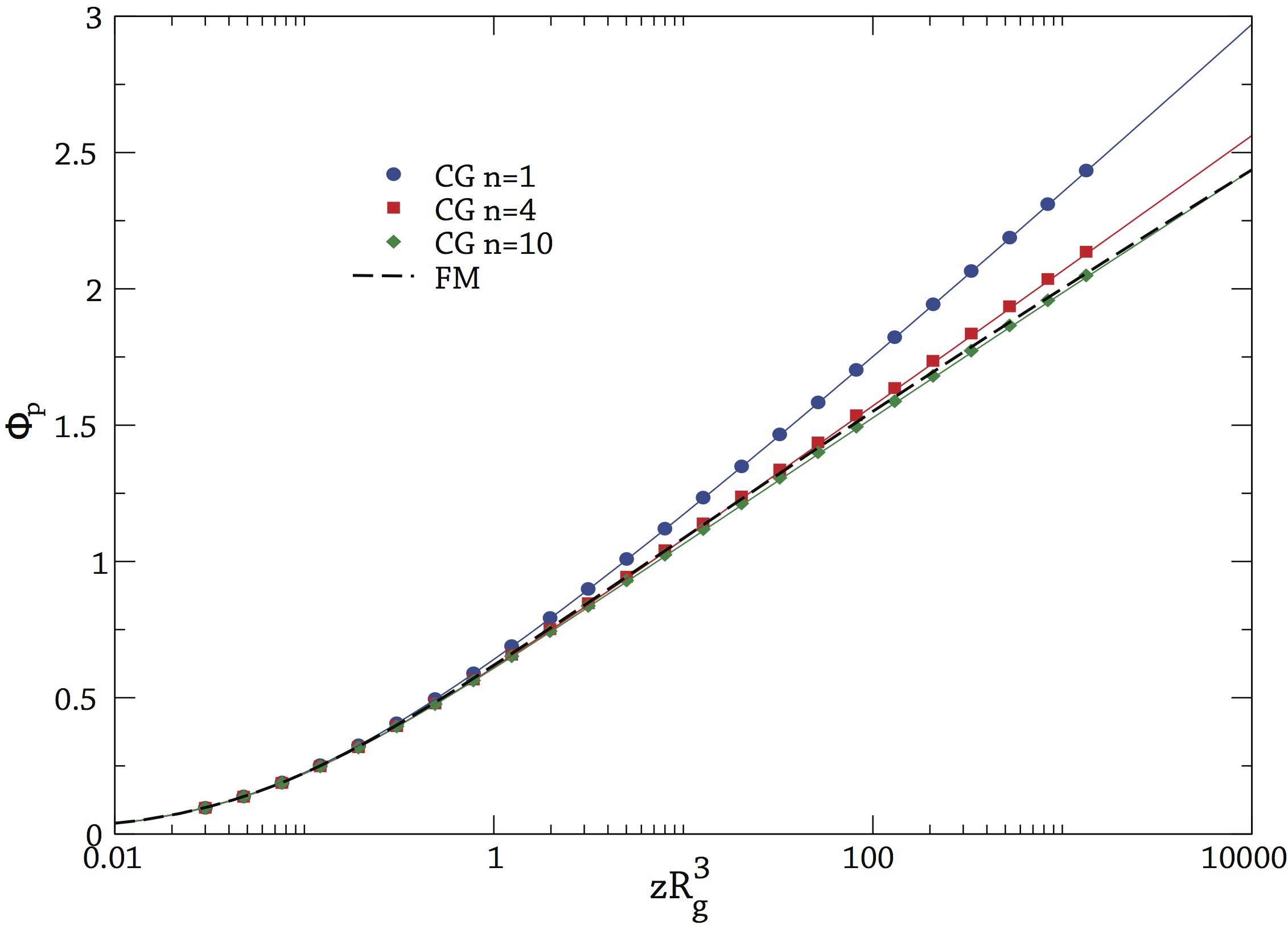} \\
\end{tabular}
\end{center}
\caption{Grand-canonical equation of state: Polymer volume fraction $\Phi_p$ 
versus $z R^3_g$, where $z$ is the fugacity defined so that $z \approx \rho_p$ 
for small densities. We report full-monomer (FM) results, and estimates obtained
by using CG models with $n=1$, 4, and 10 sites.}
\label{fig:Multiblob}
\end{figure}

In a subsequent work \cite{DPP-12-MULTI}, the tetramer model has been improved
by an additional four-body contribution which was found to be necessary to
successfully use the tetramer model as the building block for CG representations
with more blobs. By employing an {\em ad hoc} 
transferability assumption, we have shown that this new tetramer model can be
successfully used to build CG representations with more blobs per chain,
which are necessary to reach larger reduced polymer densities in the
semi-dilute regime. It was shown that this "multiblob" model reproduces the
leading correlation not explicitly considered in deriving the force field,
namely the 5-body correlation between two subsequent dihedral angles, within
$\sim$5\% of accuracy, and, more importantly, that this error does not
accumulate when increasing the number of blobs. The result is a consistent and
transferrable CG model with variable number of interaction sites per chain that
can be used to explore the thermodynamics and the structure of polymer 
solutions under good solvent conditions. For instance,
by employing a 30-blob chains, accurate results are obtained up to
$\Phi_p\simeq 8$. In this strategy, all potentials are derived at zero density 
so that standard statistical mechanics can be applied. As an example,
 we report in Fig.~\ref{fig:Multiblob} the equation of state obtained 
by grancanonical Monte Carlo simulations. 
Results for the single-site model, the tetramer model, and the decamer model 
are compared with the universal equation of state obtained at the full-monomer 
level in the scaling limit. Clearly, the accuracy improves systematically by increasing the number of sites per chain. 

The same strategy has been subsequently extended to the thermal crossover 
towards the $\theta$ point \cite{DPP-13-thermal}. In this case,
the temperature variable is expressed in terms of the parameter $z$ 
(see section 2.2). As before, we have developed the tetramer model at selected 
values of $z$. We have found that the tetramer model is  
accurate in a wider range of reduced polymer density when approaching the 
$\theta$ temperature. The transferability assumption to built multi-sites 
CG models is now more elaborated, since it must combine density and 
temperature, but it is equally successful, see Ref.~\cite{DPP-13-thermal} for details. 

More recently, we have extended the tetramer model and its $n$-mer extensions 
to colloid-polymer mixtures in a common good solvent \cite{DPP-14-colloidi}. As before we limit our intermolecular force field to pairwise central potentials. 
This strategy is successful only if many-body contributions to the free energy 
can be neglected.
For this to happen we need the blob radius of gyration to be smaller than 
the average distance between the surface of two nearby colloids. 
At given reduced colloidal density $\Phi_c$ the average radius $R_s$ of 
the sphere containing a single colloid is $R_s/R_c=\Phi_c^{-1/3}$, 
and the average distance between the surface of two nearby colloids is 
$d/R_c=2(R_s-R_c)/R_c=2(1-\Phi_c^{1/3})/\Phi_c^{1/3}$. Therefore,
 the condition is $R_b/R_c=q n^{-\nu}<2(1-\Phi_c^{1/3})/\Phi_c^{1/3}$ or,
in other terms,
$
n>q^{1/\nu}\left[ \frac{\Phi_c^{1/3}}{2(1-\Phi_c^{1/3})}\right]^{1/\nu}
$ 
which expresses the minimum number of blobs needed for given reduce colloid density and size.
Moreover, in order to treat a blob as a single site we need that $R_b/R_c<1$ 
(colloidal limit) which implies $n>q^{1/\nu}$.
Therefore, the global condition on the suitable number of blobs is
 \begin{equation}
n>q^{1/\nu}\max\left[1, \left(\frac{\Phi_c^{1/3}}{2(1-\Phi_c^{1/3})}\right)^{1/\nu} \right]
\label{eq:limit}
\end{equation}  
The term in square brackets is 1 for $\Phi_c \leq 0.3$, while is an increasing function of $\Phi_c$ for larger values.
In ref. \cite{DPP-14-colloidi} the homogenous phase of the 
mixture at various values of $q$ was investigated.  
For $q = 0.5$ and $q = 1$, tetramer and full-monomer are found to be in full agreement up to $\Phi_p= 2$. For $q = 2$, the tetramer slightly underestimates 
the depletion thickness. Nonetheless, it represents a significant improvement with respect to the single-blob model, which becomes increasingly inaccurate as $\Phi_p$ increases. Also the homogeneous phase of the mixture for increasing colloid density was studied  \cite{DPP-14-colloidi} and again the tetramer representation was found to be accurate for $q=0.5, 1$ at all values of $\Phi_p, \Phi_c$ in the single phase regions. At $q=2$, discrepancies between the full-monomer predictions and the tetramer results increase 
with increasing both $\Phi_p$ and $\Phi_c$, indicating the need of a more
refined representation. By a simple transferability assumption for the
blob-colloid potential we showed that the decamer representation can 
describe very accurately the interfacial 
properties of the system in the single-phase region.

\section{Conclusions and outlook} \label{sec:6}

In this paper we have revisited a coarse-graining strategy for polymer systems,
in which a polymer chain is reduced to a single interaction site by tracing out the intramolecular degrees of freedom. The effective potential among the CG sites is inherently many-body and can be reduced to a sum of pairwise central 
contributions either in the 
low polymer-concentration limit or by allowing this effective pair-potential to depends on the thermodynamic state of the system. 
This CG strategy has been widely used in the past 
\cite{LLWAJAR-98,WLL-99,LBHM-00,JDLFL-01,BLHM-01,Dzubiella-etal-01,%BL-02,BLH-02,DLL-02,KHL-03,AJH-04,%%
HG-04,PH-05,VJDL-05,PH-06,FBD-08,NLMC-10,DPP-12-compressible,%%
DPP-13-depletion,DPP-14-colloidi}
to study polymer solutions in their homogeneous liquid phase and to address 
the theoretical study of the phase diagram of mixtures of non-absorbing 
colloids and chains of different architectures in a solvent. 
For homogenous polymer solutions we have shown the limits of 
validity of the CG single-site model with state-independent interactions 
(derived at zero polymer density) and discussed the apparent improvement 
obtained by switching to density-dependent pair interactions. 
The latter model is indeed tuned to represent the 
pair correlation function at any finite polymer concentration but it requires 
the knowledge of such correlation for the underlying microscopic model, 
a task that need to be accomplished by simulating the microscopic model itself.
This fact points to a limited predictive character and weakens the relevance
of this strategy. Moreover, state-dependent interactions need to be used with 
care as standard thermodynamic relations do not hold. Equivalent routes to 
physical properties for state-independent potentials provide different results 
when the interaction itself depends on the thermodynamic state of the system
\cite{Louis-02,DPP-13-state-dep}.  
We have explicitly discussed the calculation of the chain chemical potential 
for the homogeneous solution in Sec.~\ref{sec:3}. 
A main consequence of this inconsistency is the failure of the equivalence 
among different statistical ensembles even in the thermodynamic limit
\cite{DPP-13-state-dep}. 

These limitation were not fully recognized at first and the state-dependent CG
model was used in grand-canonical simulations to study the demixing transition
in colloid-polymer dispersions. This CG model exhibits a demixing transition in
qualitative agreement with experiments and phenomenological theories (GFVT). 
As discussed in Sec.~\ref{sec:4},
we can now see that this agreement was accidental, since the phase transition 
in this model is driven by density fluctuations quite larger than in the
underlying microscopic system. On the other hand,
the CG model with zero-density interactions, which is thermodynamically
consistent, is quite more accurate than the other model, but its accuracy is limited to a small range of polymer densities in the homogenous phase. 
Therefore, an accurate single-site CG model to describe the demixing transition of 
colloid-polymer dispersions seems to be out of reach. 

As we have briefly reviewed in the paper, in recent years
\cite{DPP-12-Soft,DPP-12-MULTI} 
we have developed a multi-site strategy, 
based on effective potentials derived at zero density, which is able to overcome these limitations and to provide thermodynamic consistent results. 
This new strategy is based on two main ingredients. Firstly, a suitable and 
transferrable representation of the intramolecular effective interaction,
which allows us to keep an accurate description of single-chain properties 
for any number of sites per chain (level of coarse-graining of the model). 
Secondly, the possibility of using state-independent intermolecular 
interactions among CG sites of two different chains, thereby neglecting 
interactions among sites belonging to three or more chains, by 
adjusting the number of blobs per chain in such a way 
that the blob size be comparable to the smallest characteristic length present 
in the system. The strategy has been so far successfully applied to 
homopolymers under good-solvent conditions \cite{DPP-12-Soft,DPP-12-MULTI} 
and in the thermal crossover region towards the $\theta$ point 
\cite{DPP-13-thermal}, and to colloid-polymer mixtures in the homogeneous 
phase \cite{DPP-14-colloidi}. We are presently using this approach to compute 
the binodal line of the colloid-polymer mixtures at the values of $q$ for which 
full monomer data are absent. These results will provide the first 
quantitative determination of the phase diagram and will allow us to discuss 
on a quantitative ground the accuracy of phenomenological theories like the 
GFVT and the character of the experimental results (solvent quality). 
Note that it is also possible to incorporate in the model additional global
variables, for instance the radius of gyration, as in Ref.~\cite{VBK-10}
(see Ref.~\cite{DPP-12-compressible} for the a discussion in the 
single-site context).

The same strategy could be applied to the most disparate situations ranging 
from stretched and/or confined chains, networks, brushes, polymer 
nanocomposites etc. The underlying principles on which our multi-site strategy 
is based are the fractal nature of polymers and their self-similarity. 
On the other hand, we expect our strategy to fail if applied to polymer melts,
 since the many-body character of the effective potential cannot be truncated 
at the lowest order by just increasing the number of sites per chain. 
Alternative strategies have been devised for polymer melts 
(see the contribution by Guenza \cite{Guenza} in the same issue). 
However, noting that polymer-density fluctuations are much smaller in polymer 
melts than in solutions it might be possible that single-site CG models 
with state-dependent interaction provide an accurate enough description 
of the microscopic system and a much reduced thermodynamic inconsistency 
with respect to solutions. 
Other interesting systems for which only heuristic multi-site approaches have been 
used so far are di-block copolymer solutions \cite{CHC-11},
grafted polymer systems \cite{CCH-12}, 
and telechelic star polymer systems \cite{CCLLB-12}. It would be interesting to benchmark the predictions of such heuristic model against our systematical and controllable strategy.

\medskip

\section{Acknowledgments}

G.D. acknowledges support from the Italian Ministry of Education Grant PRIN
2010HXAW77. Computer time has been provided by the Pisa INFN Computer Center and 
CINECA through the ISCRA PHCOPY HP10CFFG8Q project.

% For example, with the option graphics use
%\resizebox{0.75\columnwidth}{!}{%
%  \includegraphics{fig1.eps} }
%
%
%\begin{thebibliography}{}
% and use \bibitem to create references.
%\bibitem{RefJ}
% Format for Journal Reference
%Author, Journal \textbf{Volume}, (year) page numbers
% Format for books
%\bibitem{RefB}
%Author, \textit{Book title} (Publisher, place year) page numbers
% etc
%\end{thebibliography}


\begin{thebibliography}{99}

\bibitem{Voth2009}
G. A. Voth, ed.,
\textit{Coarse-Graining of Condensed Phases and Biomolecular Systems}
(CRC Press, Boca Raton, 2009).

\bibitem{PCCP2009}
R. Feller, Guest Editor, Phys. Chem. Chem. Phys. \textbf{11} (2009) 1853

\bibitem{SM2009}
M. Wilson, Guest Editor, Soft Matter \textbf{5} (2009) 4341

\bibitem{FarDisc2010}
\textit{Multiscale Modelling of Soft Matter},
Faraday Discussion \textbf{144} (2010) 1

\bibitem{RS-67}
G. S. Rushbrooke and M. Silbert, 
Mol. Phys. {\bf 12} (1967) 505

\bibitem{Rowlinson-67}
J. S. Rowlinson, Mol. Phys. {\bf 12} (1967) 513

\bibitem{Barker}
J. A. Barker, D. Henderson, and W. R. Smith,
Mol. Phys. \textbf{17} (1969) 579

\bibitem{Casanova}
G. Casanova, R. J. Dulla, D. A. Jonah, J. S. Rowlinson,
and G. Saville, Mol. Phys. {\bf 18} (1970) 589

\bibitem{VanderHoef}
M. A. van der Hoef and P. A. Madden, 
J. Chem. Phys. \textbf{111} (1999) 1520

\bibitem{Likos-01}
C. N. Likos, Phys. Rep. {\bf 348} (2001) 267 \\
C. N. Likos, Soft Matter {\bf 2} (2006) 478

\bibitem{MullerPlathe-02}
F. M{\"u}ller-Plathe, Chem. Phys. Chem. {\bf 3} (2002) 754

\bibitem{PK-09}
C. Peter and K. Kremer, Soft Matter {\bf 5} (2009) 4357

\bibitem{HMD-06}
J. P. Hansen and I. McDonald,
{\em Theory of Simple Liquids\/}, 3rd ed.
(Academic Press, Amsterdam, 2006).

% Integral-equation theory of the structure of polymer melts
\bibitem{SC-87}
K. S. Schweizer and J. G. Curro,
Phys. Rev. Lett. {\bf 58} (1987) 246 

%% PRISM theory of the structure, thermodynamics, and phase transitions
%% of polymer liquids and alloys}, p. 319--377
\bibitem{SC-94}
K. S. Schweizer and J. G. Curro, in 
{\em Atomistic Modeling of Physical Properties}
(Springer, Berlin, 1994), p. 319.

%% Integral equation theories of the structure, thermodynamics, and phase
%% transitions of polymer fluids, pp. 1--142
\bibitem{SC-97}
K. S. Schweizer and J. G. Curro, 
Adv. Chem. Phys. {\bf 98} (1997) 1

\bibitem{DPP-13-depletion}
G. D'Adamo, A. Pelissetto, and C. Pierleoni,
Mol. Phys. {\bf 111} (2013) 3372

\bibitem{DPP-14-GFVT}
G. D'Adamo, A. Pelissetto, and C. Pierleoni,
J. Chem. Phys. {\bf 141} (2014) 024902 

% Multiscale simulation of soft matter: From scale bridging to adaptive
%resolution, pp. 545--571
\bibitem{PDSK-08}
M. Praprotnik, L. Delle Site, and K. Kremer, 
Ann. Rev. Phys. Chem. {\bf 59} (2008) 545

% Hamiltonian adaptive resolution simulation for molecular liquids
\bibitem{PFEDBKED-13}
R. Potestio, S. Fritsch, P. Espa{\~n}ol, R. Delgado-Buscalioni, K. Kremer, 
R. Everaers, and D. Donadio,
Phys. Rev. Lett. {\bf 110} (2013) 108301

%% Statistical mechanics of dilute polymer solutions. II
\bibitem{FK-54} 
P.~J. Flory and W.~R. Krigbaum, J. Chem. Phys. {\bf 18} (1950) 1086

\bibitem{GKK-82}
A. Grosberg, P. Khalatur, and A. Khokhlov, 
Makromol. Chem. Rapid Commun {\bf 3} (1982) 709

\bibitem{KS-89}
A.~B.~Kr{\" u}ger and L.~Sch{\" a}fer, 
J. Physique {\bf 50} (1989) 3191

\bibitem{DH-94}
J. Dautenhahn and C. Hall, Macromolecules {\bf 27} (1994) 5399

\bibitem{LLWAJAR-98}
C. N. Likos, H. L\"owen, M. Watzlawek, B. Abbas,
O. Jucknischke, J. Allgaier, and D. Richter,
Phys. Rev. Lett. {\bf 80} (1998) 4450

\bibitem{WLL-99}
M. Watzlawek, C. N. Likos, and H. L\"owen,
Phys. Rev. Lett. {\bf 82} (1999) 5289

% Can polymer coils be modeled as "soft colloids"?
\bibitem{LBHM-00}
A. A. Louis, P. G. Bolhuis, J. P. Hansen, and E. J. Meijer,
Phys. Rev. Lett. {\bf 85} (2000) 2522

\bibitem{JDLFL-01}
A. Jusufi, J. Dzubiella, C. N. Likos, C. von Ferber, and H. L\"owen,
J. Phys.: Condens. Matter {\bf 13} (2001) 6177

% Accurate effective pair potentials for polymer solutions
\bibitem{BLHM-01}
P. G. Bolhuis, A. A. Louis, J. P. Hansen, and E. J. Meijer,
J. Chem. Phys. {\bf 114} (2001) 4296

\bibitem{AO-54}
S. Asakura and F. Oosawa, J. Chem. Phys. {\bf 22} (1954) 1255

% Many-body interactions and correlations in coarse-grained descriptions of
% polymer solutions
\bibitem{BLH-01}
P. G. Bolhuis, A. A. Louis, and J. P. Hansen,
Phys. Rev. E {\bf 64} (2001) 021801

\bibitem{PH-05}
A. Pelissetto and J.-P. Hansen,
J. Chem. Phys. {\bf 122} (2005) 134904

\bibitem{DPP-12-Soft}
G. D'Adamo, A. Pelissetto, and C. Pierleoni,
Soft Matter {\bf 8} (2012) 5151

\bibitem{DPP-12-compressible}
G. D'Adamo, A. Pelissetto, and C. Pierleoni,
J. Chem. Phys. {\bf 136} (2012) 224905

\bibitem{RPMP-03}
D. Reith, M. P\"utz, and F. M\"uller-Plathe,
J. Comput. Chem. {\bf 24} (2003) 1624

\bibitem{LL-95}
A. Lyubartsev and A. Laaksonen,
Phys. Rev. E {\bf 52} (1995) 3730

\bibitem{Soper-96}
A. Soper, Chem. Phys. {\bf 202} (1996) 295

% How to derive and parameterize effective potentials in colloid-polymer
% mixtures
\bibitem{BL-02}
P. G. Bolhuis and A. A. Louis,
Macromolecules {\bf 35} (2002) 1860

%% Predicting the thermodynamics by using state-dependent interactions
\bibitem{DPP-13-state-dep}
G. D'Adamo, A. Pelissetto, and C. Pierleoni,
J. Chem. Phys. {\bf 138} (2013) 234107

\bibitem{Louis-02}
A. A. Louis, J. Phys.: Condens. Matter {\bf 14} (2002) 9187

\bibitem{DLL-02}
J. Dzubiella, C. N. Likos, and H. L\"owen,
J. Chem. Phys. {\bf 116} (2002) 9518  

\bibitem{MP-13}
R. Menichetti and A. Pelissetto,
J. Chem. Phys. {\bf 138} (2013) 124902

%% Coarse grain forces in star polymer melts
\bibitem{LOB-14}
L. Liu, W. K. Den Otter, and W. J Briels,
Soft Matter {\bf 39} (2014) 7874

\bibitem{DPP-13-thermal}
G. D'Adamo, A. Pelissetto, and C. Pierleoni,
J. Chem. Phys. {\bf 139} (2013) 034901

\bibitem{Vrij-76}
A. Vrij, Pure and Appl. Chem. {\bf 48} (1976) 471

%% comportamento a piccole distanze del potenziale 
\bibitem{WP-86}
T. A. Witten and P. A. Pincus,
Macromolecules {\bf 19} (1986) 2509

%% pair potential
\bibitem{HG-04}
H.-P. Hsu and P. Grassberger,
Europhys. Lett. {\bf 66} (2004) 874

\bibitem{Pelissetto-12}
A. Pelissetto, Phys. Rev. E {\bf 85} (2012) 021803

\bibitem{Attard-02}
P. Attard, {\em Thermodynamics and Statistical Mechanics: Equilibrium by
  Entropy Maximisation} (Academic Press, Waltham, Massachusetts, 2002).

\bibitem{AW-14}
D. J. Ashton and N. B. Wilding,
J. Chem. Phys. {\bf 140} (2014) 244118

\bibitem{deGennes-79}
P. G.~de Gennes, {\em Scaling Concepts in Polymer Physics} \/
(Cornell University Press, Ithaca, NY, 1979).

\bibitem{dCJ-book}
J. des Cloizeaux and G. Jannink,
{\em Polymers in Solution: Their Modelling and Structure} \/
(Clarendon, Oxford, 1990).

\bibitem{Schaefer-99}
L. Sch\"afer, {\em Excluded Volume Effects in Polymer Solutions} \/
(Springer Verlag, Berlin, 1999).

\bibitem{DJ-72}
C. Domb and G. S. Joyce, J. Phys. C {\bf 5} (1972) 956

\bibitem{CMP-06}
S. Caracciolo, B.~M. Mognetti, and A. Pelissetto, 
J. Chem. Phys. {\bf 125} (2006) 094903

\bibitem{Pelissetto-08}
A. Pelissetto, J. Chem. Phys. {\bf 129} (2008) 044901

\bibitem{RP-13}
F. Randisi and A. Pelissetto, 
J. Chem. Phys. {\bf 139} (2013) 154902.

\bibitem{MS-88}
N. Madras and A.~D. Sokal, J. Stat. Phys. {\bf 50} (1988) 109

\bibitem{Kennedy-02}
T. Kennedy, J. Stat. Phys. {\bf 106} (2002) 407

\bibitem{Clisby-10}
N. Clisby, J. Stat. Phys. {\bf 140} (2010) 349

%% comportamento a piccole distanze del potenziale 
\bibitem{vonFerber-etal-2} 
C. von Ferber, A. Jusufi, M. Watzlawek, C. N. Likos, and
H. L\"owen, Phys. Rev. E {\bf 62} (2000) 6949

\bibitem{ZSF-53}
B. H.~Zimm, W. H.~Stockmayer, and M.~Fixman,
J. Chem. Phys. {\bf 21} (1953) 1716

\bibitem{CMP-08}
S. Caracciolo, B. M. Mognetti, and A. Pelissetto,
J. Chem. Phys. {\bf 128} (2008) 065104

\bibitem{PS-suppl}
G. C. Berry, J. Chem. Phys. \textbf{44} (1966) 4550

\bibitem{PCP-suppl}
T. Norisuye, K. Kawahara, A. Teramoto, and H. Fujita,
J. Chem. Phys  \textbf{49} (1968) 4330

\bibitem{PIB-suppl}
T. Matsumoto, N. Nishioka, and H. Fujita, J. Polym. Sci.: Part A-2
Polym. Phys. \textbf{10} (1972) 23 

\bibitem{NNT-91-suppl}
Y. Nakamura, T. Norisuye, and A. Teramoto,
Macromolecules {\bf 24} (1991) 4904

\bibitem{ANNT-94-suppl}
K. Akasaka, Y. Nakamura, T. Norisuye, and A. Teramoto,
Polym. J. {\bf 26} (1994) 363

%% Theta point results 

\bibitem{KHL-03}
V. Krakoviack, J. P. Hansen, and A. A. Louis, 
Phys. Rev. E {\bf 67} (2003) 041801

\bibitem{AJH-04}
C. I. Addison, A. A. Louis, and J. P. Hansen, 
J. Chem. Phys. {\bf 121} (2004) 9612

\bibitem{SST-02}
F. H. Stillinger, H. Sakai, and S. Torquato,
J. Chem. Phys. {\bf 117} (2002) 288

\bibitem{JHGL-07}
M. E. Johnson, T. Head-Gordon, and A. A. Louis,
J. Chem. Phys. \textbf{126} (2007) 144509

\bibitem{Henderson-74}
L. Henderson, Phys. Lett. {\bf 49A} (1974) 197

% The inverse problem in classical statistical mechanics
\bibitem{CCL-84}
J. T. Chayes, L. Chayes, and E. H. Lieb, 
Comm. Math. Phys. {\bf 93} (1984) 57

\bibitem{Morita-60}
T. Morita, Prog. Theor. Phys. {\bf 23 } (1960) 829

\bibitem{Attard-91}
P. Attard, J. Chem. Phys. {\bf 94} (1991) 2370

\bibitem{Widom-63}
B. Widom, J. Chem. Phys. {\bf 39} (1963) 2802

\bibitem{FBD-08}
A. Fortini, P. G. Bolhuis, and M Dijkstra,
J. Chem. Phys. {\bf 128} (2008) 024904

\bibitem{Poon-02}
W. C. K. Poon,
J. Phys.: Condensed Matter {\bf 14} (2002) R859

\bibitem{FS-02}
M. Fuchs and K. S. Schweizer,
J. Phys.: Condensed Matter {\bf 14} (2002) R239

\bibitem{TRK-03}
R. Tuinier, J. Rieger, and C. G. de Kruif,
Adv. Colloid Interface Sci. {\bf 103} (2003) 1

\bibitem{MvDE-07}
K. J. Mutch, J. S. van Duijneveldt, and J. Eastoe,
Soft Matter {\bf 3} (2007) 155

\bibitem{FT-08}
G. J. Fleer and R. Tuinier,
Adv. Coll. Interface Sci. {\bf 143} (2008) 1

\bibitem{ME-09}
O. Myakonkaya and J. Eastoe,
Adv. Coll. Interface Sci. {\bf 149} (2009) 39

\bibitem{BLH-02}
P. G. Bolhuis, A. A. Louis, and J. P. Hansen,
Phys. Rev. Lett. {\bf 89} (2002) 128302

\bibitem{Dzubiella-etal-01}
J. Dzubiella, A. Jusufi, C. N. Likos, C. von Ferber, H. L\"owen,
J. Stellbrink, J. Allgaier, D. Richter, A. B. Schofield, P. A. Smith,
W. C. K. Poon, and P. N. Pusey,
Phys. Rev. E {\bf 64} (2001) 010401(R)

\bibitem{SDB-03}
M. Schmidt, A. R. Denton, and J. M. Brader,
J. Chem. Phys. {\bf 118} (2003) 1541

\bibitem{VJDL-05}
R. L. C. Vink, A. Jusufi, J. Dzubiella, and C. N. Likos,
Phys. Rev. E {\bf 72}s (2005) 030401(R)

\bibitem{PH-06}
A. Pelissetto and J. P. Hansen,
Macromolecules {\bf 39} (2006) 9571

\bibitem{ZVBHV-09}
J. Zausch, P. Virnau, K. Binder, J. Horbach, R. L. C. Vink,
J. Chem. Phys. {\bf 130} (2009) 064906 \\
J. Zausch, J. Horbach, P. Virnau, and K. Binder,
J. Phys.: Condens. Matter {\bf 22} (2010) 104120

\bibitem{AP-12}
M. A. Annunziata and A. Pelissetto,
Phys. Rev. E {\bf 86} (2012) 041804

\bibitem{DPP-14-colloidi}
G. D'Adamo, A. Pelissetto, and C. Pierleoni,
J. Chem. Phys. {\bf 141} (2014) 244905.

\bibitem{CVPR-06}
C.-Y. Chou, T. T. M. Vo, A. Z. Panagiotopoulos, and M. Robert,
Physica A {\bf 369} (2006) 275

\bibitem{MLP-12}
N. A. Mahynski, T. Lafitte, and A. Z. Panagiotopoulos,
Phys. Rev. E {\bf 85} (2012) 051402

\bibitem{MIP-13}
N. A. Mahynski, B. Irick, and A. Z. Panagiotopoulos,
Phys. Rev. E {\bf 87} (2013) 022309

\bibitem{ATL-02}
D. G. L. Aarts, R. Tuinier, and H. N. W. Lekkerkerker,
J. Phys.: Condens. Matt. {\bf 14} (2002) 7551.

\bibitem{FT-07}
G. J. Fleer and R. Tuinier,
Phys. Rev. E {\bf 76} (2007) 041802

\bibitem{TSPEALF-08}
R. Tuinier, P. A. Smith, W. C. K. Poon, S. U. Egelhaaf, D. G. A. L. Aarts,
H. N. W. Lekkerkerker, and G. J. Fleer,
Europhys. Lett. {\bf 82} (2008) 68002

\bibitem{LT-11}
H. N. W. Lekkerkerker and R. Tuinier,
{\em Colloids and the Depletion Interaction},
Lect. Notes Phys. {\bf 833} (Springer, Berlin, 2011)

\bibitem{TV-77}
G. M. Torrie and J. P. Valleau, 
J. Comp. Phys. 23 (1977) 197

\bibitem{PRT-14}
A. Pelissetto and F. Ricci-Tersenghi,
in {\it Large Deviations in Physics: The Legacy of the Law of Large
Numbers},
edited by A. Vulpiani, F. Cecconi, M. Cencini, A. Puglisi, and D. Vergni,
Lect. Notes Phys. {\bf 885} (2014) 161.

\bibitem{VH-04}
R. L. C. Vink and J. Horbach, 
J. Chem. Phys. {\bf 121} (2004) 3253

\bibitem{Vink-04}
R. L. C. Vink, in 
{\em Computer Simulation Studies in Condensed Matter Physics XVIII}, 
edited by D. P. Landau, S. P. Lewis, and H. B. Schuettler 
(Springer, Berlin, 2004).

%%Optimized monte carlo data analysis
\bibitem{FS-89}
A. M. Ferrenberg and R. H. Swendsen,
Phys. Rev. Lett. {\bf 63} (1989) 1195

%% Simulation studies of fluid critical behaviour
\bibitem{Wilding-97}
N. B. Wilding, J. Phys.: Condens. Matter {\bf 9} (1997) 585

%% Probability distribution of the order parameter for the three-dimensional Ising-model
%% universality class: A high-precision Monte Carlo study,
\bibitem{TB-00}
M. M. Tsypin and H. W. J. Bl\"ote,
Phys. Rev. E {\bf 62} (2000) 73

\bibitem{PCH-07-multi}
C. Pierleoni, B. Capone and J. P. Hansen, 
J. Chem. Phys. {\bf 127} (2007) 171102

\bibitem{P-09-multi}
A. Pelissetto, J. Phys.: Condens. Matter
 {\bf 21} (2009) 115108

\bibitem{DPP-12-MULTI}
G. D'Adamo, A. Pelissetto, and C. Pierleoni,
J. Chem. Phys. {\bf 137} (2012) 024901

%%Influence of topology on effective potentials: coarse-graining ring polymers
\bibitem{NLMC-10}
A. Narros, C. N. Likos, A. J. Moreno, and B. Capone,
Soft matter {\bf 10} (2010) 9601

\bibitem{VBK-10}
T. Vettorel, G. Besold, and K. Kremer, Soft Matter {\bf 6} (2010) 2282

\bibitem{Guenza}
M. G. Guenza, this volume. \\
A. J. Clark and M. G. Guenza, J. Chem. Phys. {\bf 132} (2010)  044902

\bibitem{CHC-11}
B. Capone, J.P. Hansen, and I. Coluzza,
J. Phys. Cond. Matter {\bf 23} (2011) 194102 

\bibitem{CCH-12}
I. Coluzza, B. Capone,
 and J.-P. Hansen, Soft Matter {\bf 7} (2011) 5255

\bibitem{CCLLB-12}
B. Capone, I. Coluzza, F. G. Lo Verso, C. N. Likos, and R. Blaak, 
Phys. Rev. Lett.  {\bf 109} (2012) 238301


\end{thebibliography}
\end{document}